\newcommand{\beq}{\begin{equation}}
\newcommand{\eeq}{\end{equation}}
\newcommand{\bqa}{\begin{eqnarray}}
\newcommand{\eqa}{\end{eqnarray}}
\documentclass[12pt,epsf]{article}
\usepackage{epsfig}
\def\square{\vcenter{\vbox{\hrule height.4pt
          \hbox{\vrule width.4pt height8pt
          \kern8pt\vrule width.4pt}\hrule height.4pt}}}

\begin{document}

\begin{flushleft}\hspace{12cm}
OHSTPY-HEP-T-98 \\ \hspace{12cm}
cond-mat/9806375 \\ \hspace{12cm}
June 1998  \\ \hspace{12cm}
revised March 1999  \\
\end{flushleft}

\vskip 10mm
\begin{center}
{\Large \bf Semiclassical Corrections to a 
	Static Bose-Einstein Condensate at Zero Temperature}

\vspace{.5in}
Jens O. Andersen and Eric Braaten 
 \\

{\it Department of Physics, The Ohio State University, Columbus, OH 43210}

\end{center}

\begin{abstract}
In the mean-field approximation, a trapped Bose-Einstein condensate
at zero temperature is described by the Gross-Pitaevskii equation
for the condensate, or equivalently, by the hydrodynamic equations 
for the number density and the current density.  These equations 
receive corrections from quantum field fluctuations around the mean field.
We calculate the semiclassical corrections to these equations 
for a general time-independent state of the condensate, extending 
previous work to include vortex states as well as the ground state.
In the Thomas-Fermi limit, the semiclassical corrections can be taken 
into account by adding a local correction term to the Gross-Pitaevskii 
equation.  At second order in the Thomas-Fermi expansion,
the semiclassical corrections can be taken into account by adding 
local correction terms to the hydrodynamic equations.  

\end{abstract}

\pagebreak
\small
\normalsize

\section{Introduction}
\label{sec:Intro}

The achievement of Bose-Einstein condensation 
in trapped atomic vapors \cite{BEC} has rekindled 
interest in the theory of nonhomogeneous interacting Bose gases.
This problem can be conveniently formulated as a problem in 
quantum field theory, but
the quantum field equations are extremely difficult to solve in general.
Fortunately, many of the basic properties of the condensates 
in existing experiments,
such as the density profile of the ground state and the 
frequencies for small amplitude collective oscillations of the condensate, 
can be described reasonably well using the mean-field approximation.
This approximation reduces the problem to solving the Gross-Pitaevskii 
equation, a classical equation for the mean field.
However, fluctuations of the quantum field around the mean field provide
corrections to mean-field predictions that 
grow as the square root of the number density of the atoms.
These corrections will become increasingly important as higher
condensate densities are achieved and as the precision of 
experimental measurements improves.  Furthermore,
there are some observables, such as the damping rates of collective
oscillations, that vanish in the mean-field approximation
and are therefore sensitive to the effects of quantum field fluctuations.
It is therefore important to understand the effects of quantum field 
fluctuations, and to be able to calculate them accurately.

The problem of the nonhomogeneous interacting Bose gas at low temperature 
was studied by Fetter in 1972 \cite{Fetter-72} using the Bogoliubov 
approximation.  In this approximation, the quantum field equation is 
linearized around the solution $\phi$ to the Gross-Pitaevskii equation.  
More elaborate approximations that provide a better
extrapolation to higher temperatures, such as the 
Hartree-Fock-Bogoliubov (HFB) approximation, were developed 
in the early 1980's \cite{G-S-L,H-S}.  They require
the self-consistent solution of a coupled system of equations 
consisting of a partial differential equation for $\phi$ 
and a linear equation for the quantum fluctuation field $\tilde\psi$,
both of which involve expectation values 
of operators quadratic in $\tilde \psi$. 
A critical analysis of the HFB approximation and other simpler 
approximations has recently been given by Griffin \cite{Griffin}.

One of the attractive features of the HFB approximation is that 
it is a conserving approximation  that is guaranteed to respect the 
conservation laws that follow from the symmetries of the field theory.
One of the problems with the HFB approximation is that 
it is not a gapless approximation.  In the case of a homogeneous gas, 
the HFB approximation gives a gap in the spectrum in violation of 
Goldstone's theorem.  
More generally, it fails to respect some of the 
consequences of the spontaneously broken symmetry of the theory.
The classification of approximation methods for a Bose gas according to 
whether they are conserving or gapless approximations was first made 
by Hohenberg and Martin \cite{Hohenberg-Martin}. 
A controlled approximation that corresponds 
to the truncation of a systematic expansion in a small parameter
is guaranteed to be gapless.  However such approximations 
are typically not conserving, because the conservation laws may be 
satisfied only up to corrections that are higher order in the 
expansion parameter.

Another problem with the HFB approximation
is that solving the HFB equations is a complicated computational problem.  
It requires solving the 
partial differential equation for $\phi$, solving 
the eigenvalue equations for all the normal modes of $\tilde \psi$,
calculating expectation values of operators quadratic in $\tilde \psi$, 
and iterating this sequence of calculations until they are self-consistent.  
The resulting solutions 
for $\phi$ and for the normal modes of $\tilde\psi$ 
contain a great deal of information.
Unfortunately, because the HFB approximation is not gapless,
some of that information is qualitatively wrong.
Furthermore, much of that information is not easily accessible
in experiments.  For example, the total contribution to the 
number density from all the normal modes is easier to measure than
the contribution from individual normal modes.

It would be worthwhile to develop an improved treatment of the 
nonhomogeneous interacting Bose gas that avoids the inconsistencies 
of the HFB approximation and is also computationally simpler.  
It was shown in Ref. \cite{Braaten-Nieto:2}
that the effects of quantum field fluctuations on the ground state 
can be taken into account by adding local correction terms to the 
partial differential equations of the mean-field approximation.
The consistency of the approach was guaranteed by using only 
controlled approximations.  
The enormous simplification compared to the HFB approximation 
comes at the expense of any direct information on the 
individual normal modes of the quantum fluctuation field. 
In this paper, we extend the results of Ref.~\cite{Braaten-Nieto:2} 
to arbitrary time-independent states of the Bose-Einstein condensate
at zero temperature, including vortex states.  We also streamline 
the derivations in Ref.~\cite{Braaten-Nieto:2} 
by using dimensional regularization
to control infrared and ultraviolet divergences.

The controlled approximations that guarantee the consistency 
of our approach are based on truncations of systematic expansions 
in two small quantities.  The first of these quantities 
is $[64\rho ({\bf r})a^3/\pi]^{1/2}$,
where $a$ is the S-wave scattering length and $\rho$ is the number density.
This quantity is a local measure of the magnitude of the 
effects of quantum field fluctuations.  
We restrict our analysis to the semiclassical approximation,
in which the expansion is truncated at first order in $\sqrt{\rho a^3}$.
The second small quantity is $\xi / R$, where 
$\xi = [16 \pi a \rho ( {\bf r})]^{-1/2}$ is the local coherence length 
and $R$ is the length scale for significant variations in $\rho({\bf r})$.  
The expansion in $\xi /R$ defines the Thomas-Fermi approximation.  
The effects of quantum field fluctuations are dominated by modes with 
wavelengths of order $\xi$.  The quantity $\xi /R$ is therefore a
measure of how local the effects of the quantum field fluctuations are.

In present experiments with Bose-Einstein condensates, the 
largest values of the semiclassical expansion parameter have been 
achieved for condensates of Na$^{23}$.  The scattering length of 
Na$^{23}$ is $a = 2.75$ nm.  In the experiment of Ref.~\cite{Stamper-Kurn}, 
the peak density 
that was achieved was approximately 
$\rho = 3 \times 10^{15}$ atoms/cm$^3$.
The peak value of the semiclassical expansion parameter is 
therefore $(64\rho a^3/\pi)^{1/2}= 0.04$. 
This is small but still large enough that semiclassical corrections 
should be observable in quantities that can be measured with high precision,
such as the collective excitation frequencies of the condensate. 
The value of the coherence length $\xi$ at the peak density is
$[16 \pi a \rho ( {\bf r})]^{-1/2}= 5\times 10^{-6}$ cm.
The transverse radius $R$ of the elongated condensate  
was roughly $3 \times 10^{-3}$ cm.
Thus the Thomas-Fermi expansion parameter $\xi/R$ at the peak density
is roughly 0.02.
The sizes of the expansion parameters can be controlled by adjusting 
the shape of the trapping potential and the number of atoms in the trap.
The number density that can be achieved is limited however by
the loss of atoms from the trap due to 3-body collisions.
The size of the expansion parameters can also be controlled by 
changing the scattering length of the atoms.  For example, 
the use of Feshbach resonances to change the scattering length 
of Na$^{23}$ atoms has already been demonstrated \cite{Inouye}.

In the mean-field approximation, the condensate $\phi$ for
a time-independent state of  a Bose-Einstein condensate in 
an external potential $V({\bf r})$ satisfies
the Gross-Pitaevskii equation:
\beq
0 \;=\;
\left(  {\hbar^2 \over 2 m} \nabla^2 
	+ \mu - V \right) \phi
	\;-\; {4 \pi \hbar^2 a \over m} |\phi|^2 \phi ,
\label{GP-phi}
\eeq
where $a$ is the S-wave scattering length and $\mu$ is the chemical potential. 
This equation can be written equivalently 
in the hydrodynamic form that consists of a pair of coupled partial 
differential equations for the number density $\rho = |\phi|^2$ and
the current density ${\bf j} = -i (\phi^* \nabla \phi - \phi \nabla \phi^*)$:
\bqa
\mu &=&
V \;+\; {4 \pi \hbar^2 a \over m} \rho 
\;+\; {\hbar^2 \over 8 m \rho^2} 
\left[ - 2 \rho \nabla^2 \rho + (\nabla \rho)^2 \right]
\;+\; {m \over 2 \rho^2} {\bf j}^2 ,
\label{GP-rho}
\\
0 &=& \nabla \cdot {\bf j}.
\label{j-cont}
\eqa

We calculate the corrections to the mean-field equations from
quantum field fluctuations using the semiclassical approximation, 
which includes all terms through first order in $\sqrt{\rho a^3}$.
We use the Thomas-Fermi expansion to express the corrections 
in terms of local quantities.
In the Thomas-Fermi limit, we find that the quantum 
field fluctuations can 
be taken into account by adding a local correction
term to the Gross-Pitaevskii equation (\ref{GP-phi}):
\beq
0 \;=\;
\left(  {\hbar^2 \over 2 m} \nabla^2 
	+ \mu - V \right) \phi
	\;-\; {4 \pi \hbar^2 a \over m} |\phi|^2 \phi 
	\left[ 1 + {5 \over 3} \left( {64 a^3 \over \pi} \right)^{1/2} 
		|\phi| \right].
\label{final-phi}
\eeq
At second order in the Thomas-Fermi expansion, 
the correction term is no longer
local when expressed in terms of $\phi$.  However there are also 
nonlocal terms in the relation between $\rho$ and $\phi$. 
We find that the nonlocal terms cancel to give a
local correction term to the hydrodynamic form of the mean-field equations.
The current density ${\bf j}$ still satisfies the continuity equation 
(\ref{j-cont}), while (\ref{GP-rho}) is replaced by
\begin{eqnarray}
\mu &=&
V \;+\; {4 \pi \hbar^2 a \over m} \rho 
\;+\; {\hbar^2 \over 8 m \rho^2} 
	\left[- 2 \rho \nabla^2 \rho + (\nabla \rho)^2 \right]
\;+\; {m \over 2 \rho^2} {\bf j}^2
\nonumber 
\\
&&  
\;+\;  \left( {64 a^3 \rho \over \pi} \right)^{1/2}
\left( {16 \pi \hbar^2 a \over 3 m} \rho 
	\;-\; {17 \hbar^2 \over 576 m \rho^2}  
 	\left[ - 4 \rho \nabla^2 \rho + (\nabla \rho)^2 \right] 
\right)\,.
\label{final-rho}
\end{eqnarray}
This equation was derived previously in Ref. \cite{Braaten-Nieto:2}
for the special case of the ground state for which ${\bf j}$ vanishes.
In this paper, we extend that derivation to a general time-independent
state with a nonvanishing current.
We find a remarkable cancellation of the $\bf j$-dependent terms 
in the semiclassical correction, so that the dependence on
the current $\bf j$ enters only through the classical terms.

We begin in Section \ref{sec:Static} by writing down the 
general equations that determine the condensate, the number density,
and the current density for a time-independent state of a
Bose-Einstein condensate in an external potential. 
In Section \ref{sec:Perturbative}, we set up a perturbative framework for 
calculating the local effects of quantum field fluctuations
when they are dominated by wavelengths comparable to the coherence length.
The semiclassical approximation is introduced in
Section \ref{sec:Semiclassical}.  The equation for the 
condensate in this approximation
is expressed in terms of $\rho$, ${\bf j}$, and expectation values involving 
the quantum fluctuation field. 
The Thomas-Fermi expansion is used in Section \ref{sec:Thomas-Fermi}
to calculate the expectation values to second order in the gradient expansion. 
Inserting these expectation values into the semiclassical equation,
we obtain our final result (\ref{final-rho}).
In Section \ref{sec:Condensate}, we deduce the corresponding equation 
for the condensate $\phi$ and show that it is logarithmicly sensitive to 
wavelengths much larger than the coherence length.
Finally, in Section \ref{sec:Con}, 
we summarize our results and discuss some possible generalizations.
Calculational details related to the evaluation of 
Feynman diagrams are collected in two appendices.
In Appendix~\ref{app:Integrals}, we give expressions for momentum integrals 
with dimensional regularization used for both an infrared 
and an ultraviolet cutoff.  
In Appendix~\ref{app:Diagrams}, we illustrate the use of Feynman diagrams
to calculate the coefficients in the Thomas-Fermi expansions of 
the expectation values of local quantum field operators.
The coefficients that appear in the equations for the mean field 
and for the hydrodynamic variables
are given in terms of dimensionally regularized integrals
in Appendix~\ref{app:Coefficients}.

\section{Static Bose-Einstein condensate}
\label{sec:Static}

The problem of the Bose-Einstein condensation of a large number $N$
of identical atoms in a trapping potential $V({\bf r})$
can be formulated in terms of a quantum field theory with a single complex 
field $\psi({\bf r},t)$.  The quantum field satisfies 
\beq
i \dot \psi \;+\; \left[ \nabla^2 + \mu - V({\bf r}) \right] \psi 
\;-\; {g \over 2} (\psi^\dagger \psi ) \psi \;=\; 0.
\label{qfe}
\eeq
For simplicity of notation, we have set $\hbar = 2m = 1$ in (\ref{qfe}).
Dimensional analysis can be used to reinsert
the appropriate factors of $\hbar$ and $2m$ at the end of the calculation.
The parameter $g$ in (\ref{qfe}) is proportional to the S-wave 
scattering length $a$ of the atom:
\beq
\label{g-def}
g \;=\; {8 \pi \hbar^2 a \over m} \;=\; 16 \pi a.
\label{g-a}
\eeq
The chemical potential $\mu$ in (\ref{qfe}) is to be adjusted so that the 
average number of atoms in the ground state 
$| 0 \rangle$ is a specified number $N$:
\beq
\label{mu-def}
\int d^3 r \; \langle 0 | \psi^{\dagger} \psi | 0 \rangle \;=\; N.
\eeq

We now consider an arbitrary time-independent state $| n \rangle$ of  
the Bose-Einstein condensate at zero temperature.  
If the potential traps the atoms 
in a simply-connected region of space, 
the time-independent states are the ground state and vortex states.
In the ground state, the condensate $\phi({\bf r})$
has a constant phase and the current density ${\bf j}({\bf r})$ vanishes.
In a vortex state, ${\bf j}$ is nonzero and the phase of $\phi$ changes by 
a multiple of 
$2 \pi$ along a curve that circumscribes the core of the vortex.
If the atoms are trapped in a region that is not simply-connected,
there can be more complicated states $| n \rangle$ with
time-independent number density and current.
We denote the number density and the current density in the state 
$| n \rangle$ by $\rho$ and ${\bf j}$, respectively:
\bqa
\label{rho-def}
\rho &=& \langle n | \psi^{\dagger} \psi | n \rangle, 
\\
\label{j-def}
{\bf j} &=& -i \; \langle n | 
	\psi^\dagger \nabla \psi 
		- \nabla \psi^{\dagger} \psi  
	| n \rangle.
\eqa
We also denote the condensate, which is the expectation value of $\psi$ 
in that state, by $\phi$:
\beq
\label{phi-def}
\phi \;=\; \langle n | \psi | n \rangle.
\eeq
The quantum fluctuation field $\tilde\psi$, which is defined by
\beq
\psi({\bf r},t) \;=\; \phi({\bf r}) \;+\; \tilde \psi({\bf r},t),
\label{psitilde}
\eeq
has a vanishing expectation value: $\langle n | \tilde\psi | n \rangle = 0$.
The quantum field equation (\ref{qfe}) can be expressed
as a classical field equation for $\phi$ coupled to a 
quantum field equation for $\tilde \psi$:
\bqa
0 &=& \left[ \nabla^2 + \mu - V \right] \phi 
\;-\; {g \over 2} \left[ |\phi|^2 \phi
	+ 2 \langle \tilde\psi^\dagger \tilde\psi \rangle \phi
	+ \langle \tilde\psi \tilde\psi \rangle \phi^*
	+ \langle \tilde\psi^\dagger \tilde\psi \tilde\psi \rangle
	\right],
\label{qfe-phi}
\\
0 &=& i \dot{\tilde\psi} 
\;+\; \left[ \nabla^2 + \mu - V \right] \tilde\psi 
\;-\; {g \over 2} 
\left[ 2 |\phi|^2 \tilde\psi
	+ \phi^2 \tilde\psi^\dagger
	+ 2 \phi \left( \tilde\psi^\dagger \tilde\psi
		- \langle \tilde\psi^\dagger \tilde\psi \rangle \right) 
\right.
\nonumber
\\
&& \left.
 	+ \phi^* \left( \tilde\psi \tilde\psi
		- \langle \tilde\psi \tilde\psi \rangle \right)
	+ \left( \tilde\psi^\dagger \tilde\psi \tilde\psi
		- \langle \tilde\psi^\dagger 
			\tilde\psi \tilde\psi \rangle \right)
\right].
\label{qfe-psitilde}
\eqa
The equation (\ref{qfe-phi}) for $\phi$ is just the expectation 
value of (\ref{qfe}).
The equation (\ref{qfe-psitilde}) for $\tilde\psi$ is obtained 
by subtracting (\ref{qfe-phi}) from (\ref{qfe}).
We have suppressed the state $| n \rangle$ in the expectation values 
that appear in (\ref{qfe-phi}) and (\ref{qfe-psitilde}).
Through the remainder of this paper, it will be understood that 
all expectation values are for the time-independent state $| n \rangle$.

The system of equations consisting of (\ref{qfe-phi}) and (\ref{qfe-psitilde})
involves coupled nonlinear equations for $\phi$ and $\tilde\psi$
and is extremely difficult to solve in general.
We must resort to some approximations to make this system of equations
more tractable.  
The consistency of an approximation will be guaranteed if it is 
controlled by a small expansion parameter.  From 
the example of the homogeneous Bose gas, it is clear that the
fractional size of the effects of quantum field fluctuations 
is measured locally by the dimensionless quantity 
$[64 {\rho({\bf r}) a^3}/\pi]^{1/2}$.
If the peak value of this quantity is less than one, 
one can use $\sqrt{\rho a^3}$ as an expansion parameter.
If we wish to describe the interactions of the atoms in terms of the 
S-wave scattering length $a$ only, then we are limited to first
order in $\sqrt{\rho a^3}$.  At second order in $\sqrt{\rho a^3}$,
there are ultraviolet divergences whose renormalization 
requires additional input parameters, including a pointlike contribution 
to the $3 \to 3$ scattering amplitude \cite{Braaten-Nieto:1, Braaten-Nieto:3}.
We will refer to the approximation in which we truncate at first
order in $\sqrt{\rho a^3}$ as the semiclassical approximation.

The semiclassical approximation is defined 
not by neglecting any particular terms in the equations (\ref{qfe-phi}) 
and (\ref{qfe-psitilde}), but rather by the truncation of the expansion 
in $\sqrt{\rho a^3}$ at first order. This approximation does allow those 
equations to be simplified, because not all the terms 
are necessary to achieve first order accuracy, 
but the  simplifications depend on the quantity being calculated.  
Our goal will be to achieve first order accuracy in
quantities like $\phi$, $\rho$, and ${\bf j}$ that
can be defined as expectation values of local operators.
In terms of Feynman diagrams, first order accuracy requires 
keeping all contributions from one-loop diagrams, 
while diagrams with 2 or more loops can be neglected.
The expectation value
$\langle \tilde\psi^\dagger \tilde\psi \tilde\psi \rangle$
receives contributions only from diagrams with 2 or more loops,
and it therefore can be omitted in (\ref{qfe-phi}).
We need to solve the quantum field equation (\ref{qfe-psitilde})  
only with enough accuracy to guarantee first order accuracy for the 
remaining expectation values  $\langle \tilde\psi^\dagger \tilde\psi \rangle$
and $\langle \tilde\psi \tilde\psi \rangle$ in (\ref{qfe-phi}).
The terms in (\ref{qfe-psitilde}) that are quadratic or cubic in $\tilde\psi$
contribute to $\langle \tilde\psi^\dagger \tilde\psi \rangle$
and $\langle \tilde\psi \tilde\psi \rangle$ only through diagrams 
with 2 or more loops.  We therefore need keep only those terms in 
(\ref{qfe-psitilde}) that are linear in $\tilde\psi$.
Thus to the accuracy required, (\ref{qfe-phi}) 
and (\ref{qfe-psitilde}) reduce to
\bqa
0 &=& \left[ \nabla^2 + \mu - V \right] \phi 
\;-\; {g \over 2} |\phi|^2 \phi
\;-\; g  \langle \tilde\psi^\dagger \tilde\psi \rangle \phi
\;-\; {g \over 2} \langle \tilde\psi \tilde\psi \rangle \phi^* ,
\label{SC-phi}
\\
0 &=& i \dot{\tilde\psi} 
\;+\; \left[ \nabla^2 + \mu - V \right] \tilde\psi 
\;-\; g |\phi|^2 \tilde\psi
\;-\; {g \over 2} \phi^2 \tilde\psi^\dagger .
\label{SC-psitilde}
\eqa
The expectation value $\langle \tilde\psi^\dagger \tilde\psi \rangle$
in (\ref{SC-phi}) is the noncondensate density, which is the correction 
to the number density from quantum field fluctuations.
The expectation value $\langle \tilde\psi \tilde\psi \rangle$
is called the anomalous density.
It should be emphasized that the equation (\ref{SC-psitilde}) 
is not designed to give the normal modes of $\tilde\psi$ to any specific
accuracy.  We need only solve the equation for  $\tilde\psi$
with sufficient accuracy
so that the expectation values in (\ref{SC-phi}) can be calculated with 
an error that is second order in $\sqrt{\rho a^3}$ 
relative to $|\phi|^2$.  Since these expectation values are dominated by 
modes with wavelengths comparable to the coherence length $\xi$,  
the equation for  $\tilde\psi$ need only be accurate for wavelengths
of order $\xi$.

We now compare the accuracy of the semiclassical equations 
(\ref{SC-phi}) and (\ref{SC-psitilde}) to the traditional 
approximations to the full equations (\ref{qfe-phi}) and (\ref{qfe-psitilde}).
While the main purpose of the traditional approximations 
was to take into account the effects of nonzero temperature,
it is instructive to determine their accuracy at zero temperature.
The Bogoliubov approximation is defined by the equations
(\ref{SC-phi}) and (\ref{SC-psitilde}), with the expectation values  
set to zero in (\ref{SC-phi}). 
Thus $\phi$ satisfies 
the Gross-Pitaevskii equation.  The solution to this equation
gives $\phi({\bf r})$ with a fractional error that 
is first order in $\sqrt{\rho a^3}$.  In contrast, the solution to 
the semiclassical equations (\ref{SC-phi}) and (\ref{SC-psitilde}) 
gives $\phi$ with a fractional error that is second
order in $\sqrt{\rho a^3}$.

In the Hartree-Fock-Bogoliubov (HFB) approximation, 
the $\langle \tilde\psi^\dagger \tilde\psi \tilde\psi \rangle$
term in (\ref{qfe-phi}) is neglected so that it reduces to (\ref{SC-phi}) 
and the products of field operators in (\ref{qfe-psitilde}) are replaced by
terms in which pairs of operators have been contracted 
into expectation values:
\beq
0 \;=\; i \dot{\tilde\psi} 
\;+\; \left[ \nabla^2 + \mu - V \right] \tilde\psi 
\;-\; g \left[ |\phi|^2 
	+ \langle \tilde\psi^\dagger \tilde\psi \rangle \right] 
	\tilde\psi
\;-\; {g \over 2} \left[ \phi^2
	+ \langle \tilde\psi \tilde\psi \rangle \right]
	\tilde\psi^\dagger .
\label{HFB}
\eeq
Solving the system of equations ({\ref{SC-phi}) and (\ref{HFB})
is a complicated numerical problem.  It requires
making an initial guess for $\phi$ and for the 
infinitely many normal modes of $\tilde\psi$,
calculating the expectation values 
$\langle \tilde\psi^\dagger \tilde\psi \rangle$
and $\langle \tilde\psi \tilde\psi \rangle$,
solving ({\ref{SC-phi}) for $\phi$, solving (\ref{HFB}) for the
normal modes of $\tilde\psi$, 
and iterating this sequence of calculations until they are self-consistent.
The net effect of the additional terms in (\ref{HFB})
that do not appear in ({\ref{SC-psitilde}) is to include a subset of
corrections that are second order or higher in $\sqrt{\rho a^3}$.
Since there are other corrections that are second order in 
$\sqrt{\rho a^3}$ that have not been included, the use of (\ref{HFB})
in conjunction with (\ref{SC-phi}) does not improve the accuracy in
the determination of $\phi$.
If all the second order corrections were uniformly small in $\sqrt{\rho a^3}$,
it would do no harm to include only some of them.  Unfortunately,
there are individual correction terms that 
diverge in the limit $\xi/R \to 0$, but which cancel when all 
terms of a given order in $\sqrt{\rho a^3}$ are added together.
Including some but not all of the second order terms is dangerous, 
because one may omit some of the terms that are necessary for the 
cancellations.
Thus, by including a subset of the higher order corrections,
the HFB approximation may actually decrease the accuracy.
This problem can be avoided by using a controlled approximation,
because any such cancellations will occur 
order by order in the expansion parameter.

The fact that the HFB approximation is not a controlled approximation
is reflected at nonzero temperature in the existence of an energy gap 
in the spectrum of the 
hamiltonian for $\tilde\psi$, which contradicts Goldstone's theorem.
In the Popov approximation, the energy gap is eliminated by 
setting the anomalous density $\langle \tilde\psi \tilde\psi \rangle$
to zero in ({\ref{SC-phi}) and (\ref{HFB}).
The resulting error in $\phi$ is first order in $\sqrt{\rho a^3}$.
Thus at zero temperature, the Popov approximation is no more accurate 
than the Bogoliubov approximation,
despite the fact that it requires much more complicated calculations.  
In the approximation used by Goldman et al. in Ref. \cite{G-S-L},
the $\langle \tilde\psi \tilde\psi \rangle$ term in 
({\ref{SC-phi}) was set to zero.  Thus it also gives an error in $\phi$ 
that is first order in $\sqrt{\rho a^3}$.

The semiclassical equations (\ref{SC-phi}) and (\ref{SC-psitilde})
can be greatly simplified by using a further approximation
that is also controlled by a small expansion parameter.
The effects of quantum field 
fluctuations are dominated by wavelengths comparable to the 
local coherence length 
$\xi= [ 16 \pi \rho({\bf r}) a]^{-1/2}$.  
For the ground state
of a Bose-Einstein condensate containing $N$ atoms,
$\xi$ scales like $N^{-1/5}$, where $N$ is the number of 
atoms in the trap   \cite{Baym-Pethick}.  In contrast, the 
length scale $R$ for significant variations in $\rho({\bf r})$ 
grows like $N^{1/5}$.
If $N$ is sufficiently large, $\xi$ is much shorter than $R$.
This justifies an expansion in powers of $\xi/R$ or, equivalently,
in powers of gradients of $\phi$.
This expansion defines the Thomas-Fermi approximation.  
We will calculate the semiclassical corrections 
through second order in the Thomas-Fermi expansion.

In the Thomas-Fermi limit, 
the semiclassical equation ({\ref{SC-phi}) for the condensate 
becomes particularly simple. 
As we shall see, the expectation values at leading order in $\xi/R$
are simply functions of $\phi$:
\bqa
\langle \tilde\psi^\dagger \tilde\psi \rangle 
	&\approx& {1 \over 24 \pi^2} g^{3/2} |\phi|^3,
\label{ave-normal}
\\
\langle \tilde\psi \tilde\psi \rangle 
	&\approx& {1 \over 8 \pi^2} g^{3/2} |\phi| \phi^2.
\label{ave-anom}
\eqa
Inserting the expectation values (\ref{ave-normal}) and (\ref{ave-anom})
into ({\ref{SC-phi}), it
reduces to a partial differential equation for $\phi$.
Using dimensional analysis to reinsert the appropriate factors of
$2m$ and $\hbar$, we obtain the equation (\ref{final-phi}), which
is simply the Gross-Pitaevskii equation with a correction term 
that corresponds to the local density approximation.

To go to second order in the Thomas-Fermi expansion,
we need to include terms in the expectation values 
(\ref{ave-normal}) and (\ref{ave-anom}) that are second order in 
gradients of $\phi$.  As we shall see, 
the coefficients of these terms are logarithmicly infrared divergent, 
indicating a sensitivity to length scales much greater 
than $\xi$.  This sensitivity can be reduced
by expressing the equation ({\ref{SC-phi}) 
in terms of the hydrodynamic variables $\rho$ and ${\bf j}$
defined by (\ref{rho-def}) and (\ref{j-def}):
\bqa
\label{rho-HF}
\rho &=& |\phi|^2 
\;+\; \langle \tilde\psi^{\dagger} \tilde\psi \rangle, 
\\
\label{j-HF}
{\bf j} &=& -i \left( \phi^* \nabla \phi 
		- \phi \nabla \phi^* \right)
\;-\; i \langle \tilde\psi^{\dagger} \nabla \tilde\psi 
		- \nabla \tilde\psi^{\dagger} \tilde\psi \rangle.
\eqa
We will find that if the condensate $\phi$ in ({\ref{SC-phi}) is eliminated 
in favor of $\rho$ and ${\bf j}$, the infrared divergences cancel
and the equation reduces to (\ref{final-rho}).
We will obtain this result in two different ways.
In Sections~\ref{sec:Semiclassical} and \ref{sec:Thomas-Fermi},
we organize the calculation so that ({\ref{SC-phi}) 
gives directly our 
final equation (\ref{final-rho}) for $\rho$ and ${\bf j}$.
In Section~\ref{sec:Condensate}, we calculate the expectation values 
(\ref{ave-normal}) and (\ref{ave-anom}) to second order in the 
Thomas-Fermi expansion using an infrared cutoff.
We then use the condensate equation ({\ref{SC-phi}) to deduce
our final equation (\ref{final-rho}) for $\rho$ and ${\bf j}$.

\section{Perturbative framework}
\label{sec:Perturbative}

In this section, we set up a perturbative framework that can be used to 
calculate the leading effects of quantum field fluctuations.
This framework involves introducing an arbitrary momentum scale
$\Lambda$ that will later be chosen to be the inverse of the local 
coherence length $\xi$.  The perturbation series is an expansion 
in powers of $g \Lambda$, and the semiclassical approximation introduced
in Section \ref{sec:Semiclassical} is the truncation of that expansion 
at first order in $g \Lambda$.

The quantum field theory associated with the equation
(\ref{qfe}) is summarized by the action
\beq 
S[\psi] \;=\;
\int dt \int d^3x \left\{
{i \over 2} \left( \psi^{\dagger} \dot{\psi} 
	- \dot{\psi}^{\dagger} \psi \right)
-\nabla\psi^{\dagger}\cdot\nabla\psi
\;+\; \left[ \mu - V({\bf r}) \right] \psi^{\dagger}\psi
\;-\; {g \over 4}(\psi^{\dagger}\psi)^2 \right\}.
\label{action}
\eeq
This action is invariant under the $U(1)$ symmetry 
in which the field is multiplied by a phase: 
\beq 
\psi({\bf r},t)  \;\longrightarrow\; e^{i \alpha} \psi({\bf r},t).
\label{U1-sym}
\eeq
The action is also invariant under time-reversal symmetry: 
\beq 
\psi({\bf r},t)  \;\longrightarrow\; \psi^\dagger({\bf r},-t).
\label{time-sym}
\eeq
The $U(1)$ symmetry (\ref{U1-sym}) guarantees that the number density 
and the current density satisfy the continuity equation:
\beq 
\dot \rho \;+\; \nabla \cdot {\bf j}  \;=\; 0.
\label{continuity}
\eeq
A $U(1)$ symmetry transformation can also be used to make the 
condensate $\phi({\bf r}) = \langle \psi({\bf r}) \rangle$
real-valued at any specific point ${\bf r}_0$.

The quantum field theory defined by (\ref{action})
has ultraviolet divergences that must be removed by 
renormalizations of the parameters $\mu$ and $g$.  
There is also an ultraviolet divergence in the number density that 
can be removed either by renormalization 
or by an operator-ordering prescription.
In Ref. \cite{Braaten-Nieto:2}, the effects of quantum field fluctuations 
on the ground state were calculated using a momentum cutoff 
to regularize the ultraviolet divergences.  The divergences 
were cancelled explicitly by counterterms for $\mu$, $g$, and $\rho$.
In this paper, we choose to  
regularize the ultraviolet divergences 
using dimensional regularization, which involves calculating 
integrals as analytic functions of the number of dimensions 
$D$ and then analytically continuing them to $D=3$.  
This method has been used to calculate the properties of a
homogeneous Bose gas \cite{Ravndal,Braaten-Nieto:3}
and it greatly streamlines the calculations.
One advantage of dimensional regularization is that it sets 
integrals that diverge like a power of the ultraviolet cutoff, 
and involve no other momentum scales, equal to zero.
Since the counterterms for
$\mu$, $g$, and $\rho$ are pure power ultraviolet divergences,
they are identically zero in dimensional regularization. 
We have therefore omitted counterterms for the parameters
$g$ and $\mu$ in (\ref{action}).  

It is convenient to decompose both the condensate (\ref{phi-def}) 
and the quantum fluctuation field $\tilde\psi$ defined by (\ref{psitilde})
into real and imaginary parts:
\bqa
\phi({\bf r}) &=& v({\bf r}) + i w({\bf r}),
\label{phi-vw}
\\
\tilde\psi({\bf r},t) 
	&=& {\xi({\bf r},t) + i \eta({\bf r},t) \over \sqrt{2}}.
\label{psitilde-xieta}
\eqa
In the neighborhood of a point ${\bf r}_0$ where $w$ vanishes,
we can identify $\xi$ and $\eta$ as the quantum fluctuation fields 
associated with the number density 
and with the phase of the condensate, respectively.
Inserting these expressions into the action
(\ref{action}) and expanding in powers
of $\xi$ and $\eta$, we obtain a condensate term $S[v + i w]$
and terms up to fourth order in $\xi$ and $\eta$.
The equations (\ref{psitilde}) and (\ref{psitilde-xieta}) define
the Cartesian parameterization of the quantum field $\psi$ 
in terms of real quantum fields.  One could equally well use an
alternative parameterization, such as the polar parameterization:
\beq
\psi({\bf r},t) \;=\; 
\left[ v^2({\bf r}) + \sigma({\bf r},t) \right]^{1/2}
e^{i \theta({\bf r})} e^{i \alpha({\bf r,t})},
\label{polar}
\eeq
where $v$ and $\theta$ are real classical fields,
while $\sigma$ and $\alpha$ are real quantum fields.
The polar parameterization has the advantage in perturbative calculations 
that it eliminates infrared divergences that with other 
parameterizations cancel only after adding Feynman diagrams
\cite{Popov}.
If we use controlled approximations based on systematic expansions 
in small quantities,
any parameterization
should lead to the same results for physical quantities.
In Ref. \cite{Braaten-Nieto:2}, it was verified explicitly that 
the Cartesian and polar parameterizations give the same 
final equation for the number density in the ground state.
Since the calculations were simpler using the Cartesian parameterization,  
we use only that parameterization in this paper.

If the system is near the Thomas-Fermi limit,
the length scale for significant changes
in the condensate is large compared to the local coherence length
$(g \rho)^{-1/2}$.
At any specific point ${\bf r}_0$ inside the condensate, 
the short-wavelength modes of the quantum fields
$\xi$ and $\eta$ behave locally like those of 
a homogeneous Bose gas with coherence length
$[g \rho({\bf r}_0)]^{-1/2}$.  Their dispersion relation
can be approximated by the Bogoliubov dispersion relation
\beq
\epsilon(k) \;=\; k \sqrt{k^2+\Lambda^2},
\label{bog}
\eeq
with $\Lambda^2 \approx g \rho({\bf r}_0)$.
A particle with the dispersion relation (\ref{bog})
is described by a free field theory with the action 
\beq
S_{\mbox{\footnotesize free}}[\xi ,\eta] \;=\;
\int dt \int d^3x \left\{
{1 \over 2} \left( \eta \dot{\xi} - \xi \dot{\eta} \right)
\;-\; {1 \over 2} \left[ (\nabla \xi)^2 + \Lambda^2 \xi^2
	+ (\nabla \eta)^2 \right]
\right\}.
\label{S-free}
\eeq
We can express  the complete action (\ref{action}) as the sum of 
the background term $S[v+iw]$, the free action (\ref{S-free})
for the quantum fields, and an interaction term:
\beq
S[\psi] \;=\;
S[v+iw] \;+\; S_{\mbox{\footnotesize free}}[\xi,\eta]
\;+\; S_{\mbox{\footnotesize int}}[v,w,\xi,\eta].
\label{S-123}
\eeq
The last term describes the interaction
of the Bogoliubov modes with each other and with sources that depend
on $v$ and $w$:
\bqa
S_{\mbox{\footnotesize int}}[v,w,\xi,\eta] 
\;=\; \int dt \int d^3x 
\left\{
	\sqrt{2}(T \xi + S \eta)
	\;+\; {1 \over 2}(X \xi^2 + Y \eta^2 + 2 R \xi \eta)
\right. 
\nonumber
\\
\left.
	\;+\; {1 \over \sqrt{2}} (Z \xi + Q \eta)(\xi^2 + \eta^2)
	\;-\; {g \over 16} (\xi^2 + \eta^2)^2 
\right\}.
\label{S-int}
\eqa
The sources are 
\bqa
T &=& \left[\mu - V({\bf r}) - {g \over 2}(v^2 + w^2)\right] v
	+ \nabla^2 v,
\\
S &=& \left[ \mu - V({\bf r}) - {g \over 2} (v^2 + w^2) \right] w
	+ \nabla^2 w,
\\
X &=& \Lambda^2 + \mu - V({\bf r}) - {g \over 2}(3v^2 + w^2),
\label{X-def}
\\
Y &=& \mu - V({\bf r}) - {g \over 2} (v^2 + 3w^2),
\label{Y-def}
\\
Z &=& - {g \over 2} v,
\\
Q &=& - {g \over 2} w,
\\
R &=& - g v w.
\label{R-def}
\eqa
The sources $T$ and $S$ are simply the real and imaginary parts of the 
Gross-Pitaevskii equation for the condensate $v + i w$.
Note that the dependence of the source $X$ on $\Lambda^2$ precisely cancels
the $\Lambda$--dependence of the free action (\ref{S-free}). 
Thus $\Lambda$ is a completely arbitrary parameter.
That arbitrariness will be exploited below to ensure that the terms in 
(\ref{S-int}) can be treated as perturbations.

The quantum field equations for $\xi$ and $\eta$ are obtained by 
varying the action (\ref{S-123}).  Taking the expectation value
of those quantum field equations and using the fact that $\xi$ and $\eta$
have vanishing expectation values, 
we obtain a pair of equations equivalent to (\ref{qfe-phi}):
\bqa
0 &=& T \;+\; {1 \over 2} Z \langle 3 \xi^2 + \eta^2 \rangle
\;+\; Q \langle \xi\eta \rangle
\;-\; {g \over 4 \sqrt{2}} \langle \xi (\xi^2 + \eta^2) \rangle,
\label{tad-xi}
\\
0 &=& S \;+\; {1 \over 2} Q \langle \xi^2 + 3 \eta^2 \rangle
\;+\; Z \langle \xi \eta \rangle
\;-\; {g \over 4 \sqrt{2}} \langle \eta (\xi^2 + \eta^2) \rangle .
\label{tad-eta}
\eqa
We will refer to these as the tadpole equations for $\xi$ and 
for $\eta$, respectively. 
The quantum field equations for $\xi$ and $\eta$ that correspond to 
(\ref{qfe-psitilde}) are 
\bqa
0 &=& \dot{\xi}  \;+\; (\nabla^2 + Y) \eta \;+\; R \xi
\;+\; {1 \over \sqrt{2}} Q 
	\left( \xi^2 + 3 \eta^2 - \langle \xi^2 + 3 \eta^2 \rangle \right)
\nonumber 
\\
&& \;+\; \sqrt{2} Z
	\left( \xi \eta - \langle \xi \eta \rangle \right)
\;-\; {g \over 4} 
	\left( \eta (\xi^2 + \eta^2) 
		- \langle \eta (\xi^2 + \eta^2) \rangle \right),
\label{qfe-xi}
\\
0 &=& - \dot{\eta} \;+\; (\nabla^2 - \Lambda^2 + X) \xi \;+\; R \eta
\;+\; {1 \over \sqrt{2}} Z 
	\left( 3 \xi^2 + \eta^2 - \langle 3 \xi^2 + \eta^2 \rangle \right)
\nonumber 
\\
&& \;+\; \sqrt{2} Q 
	\left( \xi \eta - \langle \xi \eta \rangle \right)
\;-\; {g \over 4} 
	\left( \xi (\xi^2 + \eta^2) 
		- \langle \xi (\xi^2 + \eta^2) \rangle \right).
\label{qfe-eta}
\eqa
The expressions (\ref{rho-HF}) and (\ref{j-HF}) for the number density and
the current density can be written as
\bqa
\rho &=& 
v^2 + w^2 + {1 \over 2} \langle \xi^2 + \eta^2 \rangle,
\label{rho-xieta}
\\
{\bf j} &=&
2 \left( v \nabla w 
	- w \nabla v \right)
\;+\; \langle \xi \nabla \eta
		- \eta \nabla \xi \rangle.
\label{j-xieta}
\eqa
Our strategy will be to solve the quantum field equations
(\ref{qfe-xi}) and (\ref{qfe-eta}), treating the terms in (\ref{S-int})
as perturbations, and then to calculate the expectation values in 
(\ref{tad-xi}), (\ref{tad-eta}), (\ref{rho-xieta}),
and (\ref{j-xieta}) to first order in $g \Lambda$.

\section{Semiclassical approximation}
\label{sec:Semiclassical}

For the perturbation theory defined by the decomposition
of the action (\ref{S-123}) into free and interaction parts,
the appropriate dimensionless expansion parameter is 
the product of the coupling constant
$g$ and the coherence scale $\Lambda$.
The semiclassical approximation is defined by truncating
the expansion in powers of $g \Lambda$ at first order.
If the dimensionless ratio $\Lambda/\sqrt{g \rho}$ is held fixed,
this is equivalent to truncating at first order in $\sqrt{\rho a^3}$.
In terms of Feynman diagrams, accuracy to first order in $g \Lambda$ 
corresponds to keeping only the contributions from one-loop diagrams.
The expectation values in (\ref{tad-xi}) and (\ref{tad-eta})
that are cubic in $\xi$ and $\eta$
receive contributions only from diagrams with 2 or more loops 
and therefore can be omitted.  Thus the tadpole equations reduce to
\bqa
\label{t1}
0 &=& T \;+\; {1 \over 2} Z \langle 3 \xi^2 + \eta^2 \rangle
\;+\; Q \langle \xi\eta \rangle,
\\
\label{t2}
0 &=& S \;+\; {1 \over 2} Q \langle \xi^2 + 3 \eta^2 \rangle
\;+\; Z \langle \xi \eta \rangle.
\eqa
We will refer to these equations as the semiclassical tadpole equations for 
$\xi$ and $\eta$, respectively.
In the quantum field equations (\ref{qfe-xi}) and (\ref{qfe-eta}),
the terms that are quadratic or cubic in $\xi$ and $\eta$ 
contribute to the expectation values in (\ref{t1}) and (\ref{t2})
only through diagrams with 2 or more loops.  We therefore need to 
keep only the linear terms in the quantum field equations:
\bqa
\label{xi-HF}
0 &=& \dot \xi \;+\; (\nabla^2 + Y) \eta \;+\; R \xi,
\\
\label{eta-HF}
0 &=& - \dot \eta \;+\; (\nabla^2 - \Lambda^2 + X) \xi \;+\; R \eta.
\eqa
These equations describe the propagation of the fields
$\xi$ and $\eta$ in the presence of the sources $X$, $Y$, and $R$,
but with no other interactions. 
Note that the explicit dependence of (\ref{eta-HF}) on $\Lambda$
is cancelled by the $\Lambda$--dependence of the source $X$.
We will find that a judicious choice of $\Lambda$ will allow the source
$X$ to be treated as a perturbation, along with $Y$ and $R$.
The expectation values that appear in
the semiclassical tadpole equations (\ref{t1}) and (\ref{t2}) 
and in the expressions for $\rho$ and ${\bf j}$
in (\ref{rho-xieta}) and (\ref{j-xieta}) are 
functionals of the sources 
$X$, $Y$, and $R$.  Under the time-reversal symmetry (\ref{time-sym}),
$\xi$, $X$, and $Y$ are even, while $\eta$ and $R$ are odd.
The expectation values $\langle \xi^2 \rangle$ and $\langle \eta^2 \rangle$
are therefore even functionals of $R$, while $\langle \xi \eta \rangle$
and $\langle \xi \nabla \eta - \eta \nabla \xi \rangle$ 
are odd functionals of $R$.  

Since the sources $X$, $Y$, and $R$ depend on $v$ and $w$,
the expectation values in (\ref{rho-xieta})--(\ref{t2})
are functionals of $v$ and $w$.
In order to obtain an equation for $\rho$ and ${\bf j}$,
we will use (\ref{rho-xieta}) and (\ref{j-xieta})
to eliminate $v$ and $w$ from (\ref{t1}) and (\ref{t2})
in favor of $\rho$ and ${\bf j}$.
The expectation values that appear in these equations 
are the corrections 
to the classical equations from quantum field fluctuations.  
They are smaller than individual terms in
the classical equations by a factor of $\sqrt{\rho a^3}$.
The semiclassical approximation includes all terms through
first order in $\sqrt{\rho a^3}$.
In the course of manipulating the equations (\ref{rho-xieta})--(\ref{t2})
to eliminate $v$ and $w$, there is no loss of accuracy if we expand
the resulting expressions to first order in the expectation values, 
dropping higher order terms. The resulting equations still
includes all effects of quantum field fluctuations through first order in 
$\sqrt{\rho a^3}$.
Solving (\ref{rho-xieta}) and (\ref{j-xieta}) for $v$ and $\nabla w$
and expanding to first order in the expectation values, we obtain
\bqa
\label{v}
v &=& (\rho -w^2)^{1/2} 
\;-\; {1 \over 4 (\rho-w^2)^{1/2}} \langle \xi^2 + \eta^2 \rangle,
\\
\label{dw}
\nabla w &=& {1 \over 2v} ({\bf j} + 2 w \nabla v)
\;-\; {1 \over 2v} \langle \xi \nabla \eta - \eta \nabla \xi \rangle.
\eqa
We can derive analogous expressions for $\nabla v$, $\nabla^2 v$, 
and $\nabla^2 w$ by differentiating (\ref{v}) and (\ref{dw}).
By substituting iteratively and expanding to first order in
the expectation values, we can eliminate 
$v$, $\nabla v$, $\nabla^2 v$, $\nabla w$, and $\nabla^2 w$
from the right side of the equation in favor of $\rho$, ${\bf j}$, 
their derivatives, and also $w$.  
The resulting expressions at a specific point 
${\bf r}_0$ become particularly simple 
if we use a global phase transformation to set $w({\bf r}_0)=0$:
\bqa
v({\bf r}_0) &=& {1 \over 4 \rho^{1/2}}
\left[ 4 \rho \;-\; \langle \xi^2 + \eta^2 \rangle \right],
\label{list1} 
\\
\nabla v({\bf r}_0) &=& {1 \over 8 \rho^{3/2}}
\left[ 4 \rho \nabla \rho
	\;+\; \nabla \rho \langle \xi^2 + \eta^2 \rangle
	\;-\; 2 \rho \nabla \langle \xi^2 + \eta^2 \rangle \right],
\\
\nabla^2 v({\bf r}_0) &=& {1 \over 16 \rho^{5/2}}
\left[ -4 \rho [ - 2 \rho \nabla^2 \rho + (\nabla \rho)^2 + {\bf j}^2 ]
	\;-\; [ - 2 \rho \nabla^2 \rho + 3 (\nabla \rho)^2 + 3 {\bf j}^2 ]
	\langle \xi^2 + \eta^2 \rangle 
\right.
\nonumber
\\
&& \left.
\;+\; 4 \rho \nabla \rho \cdot 
	\nabla \langle \xi^2 + \eta^2 \rangle
\;-\; 4 \rho^2 \nabla^2 \langle \xi^2 + \eta^2 \rangle
\;+\; 8 \rho {\bf j} \cdot
	\langle \xi \nabla \eta - \eta \nabla \xi \rangle
\right] ,
\\
w({\bf r}_0) &=& 0, 
\\
\nabla w({\bf r}_0) &=& {1 \over 8 \rho^{3/2}} 
\left[ 4 \rho {\bf j}
	\;+\; {\bf j} \langle \xi^2 + \eta^2 \rangle
	\;-\; 4 \rho \langle \xi \nabla \eta - \eta \nabla \xi \rangle \right],
\\
\nabla^2w({\bf r}_0) &=& {1 \over 8 \rho^{3/2}}
\left[ 4 \rho \nabla \cdot {\bf j}
	\;+\; \nabla \cdot {\bf j} \langle \xi^2 + \eta^2 \rangle
	\;-\; 4 \rho \nabla \cdot 
		\langle \xi \nabla \eta - \eta \nabla \xi \rangle 
\right].
\label{list6}
\eqa

Inserting the expressions (\ref{list1})--(\ref{list6})
into (\ref{t1}) and (\ref{t2}) and expanding to first order in 
the expectation values, we obtain expressions for the tadpole equations
at the point ${\bf r}_0$ with $v$ and $w$ eliminated in favor of 
$\rho$ and ${\bf j}$.  The semiclassical
tadpole equation (\ref{t2}) for $\eta$ reads
\beq
0 \;=\; {1 \over 8 \rho^{3/2}} 
\left[ 4 \rho \nabla \cdot {\bf j}
	\;+\; \nabla \cdot {\bf j} \, 
		\langle \xi^2 + \eta^2 \rangle
	\;-\; 4 g \rho^2 \, \langle \xi \eta \rangle
	\;-\; 4 \rho \nabla \cdot 
		\langle \xi \nabla \eta - \eta \nabla \xi \rangle 
\right].
\label{tadpole-eta}
\eeq
The last term in (\ref{tadpole-eta}) 
can be simplified by using the quantum field
equations (\ref{xi-HF}) and (\ref{eta-HF}) for $\xi$ and $\eta$:
\bqa
\nabla \cdot \langle \xi \nabla \eta - \eta \nabla \xi \rangle
&=& \langle \xi \nabla^2 \eta - \eta \nabla^2 \xi \rangle
\nonumber
\\
&=& - {1 \over 2} {d \ \over dt} \langle \xi^2 + \eta^2 \rangle
	\;+\; (X - Y - \Lambda^2) \langle \xi \eta \rangle
	\;-\; R \langle \xi^2 - \eta^2 \rangle
\nonumber
\\
&=& - g (v^2 - w^2) \langle \xi \eta \rangle
\;-\; g v w \langle \xi^2 - \eta^2 \rangle.
\label{simplify}
\eqa
In the last step, we have used the expressions (\ref{X-def}),
(\ref{Y-def}), and (\ref{R-def}) for the sources and the fact that 
$\langle \xi^2 + \eta^2 \rangle$ is time-independent.
Since $w$ vanishes at the point ${\bf r}_0$ and $\rho - v^2$ is of order 
$\sqrt{\rho a^3}$ at that point, (\ref{simplify}) reduces to 
$- g \rho  \langle \xi \eta \rangle$, up to terms that are second order 
in $\sqrt{\rho a^3}$.  Thus
the last two terms in (\ref{tadpole-eta}) cancel.
Up to an overall multiplicative factor, 
the first two terms in (\ref{tadpole-eta}) reduce to the
time-independent continuity equation $\nabla \cdot {\bf j} = 0$.

The semiclassical tadpole equation (\ref{t1}) for $\xi$  
at the point ${\bf r}_0$ reduces to
\bqa
0 &=& 
\left( \mu - V - {g \over 2} \rho
	- {1 \over 4 \rho^2} 
	\left[ - 2 \rho \nabla^2 \rho  + (\nabla \rho)^2 + {\bf j}^2 \right] 
\right) \rho^{1/2} 
\nonumber
\\ 
&& \;-\; {g \over 2} \rho^{1/2} \, \langle \xi^2 \rangle
\;-\; {1 \over 4 \rho^{5/2}}
	\left[ - \rho \nabla^2 \rho + (\nabla \rho)^2 + {\bf j}^2 \right]
	\langle \xi^2 + \eta^2 \rangle
\nonumber
\\ 
&&
\;+\; {1 \over 4 \rho^{3/2}} \nabla \rho \cdot 
	\nabla \langle \xi^2 + \eta^2 \rangle
\;-\; {1 \over 4 \rho^{1/2}} \nabla^2 \langle \xi^2 + \eta^2 \rangle
\;+\; {1 \over 2 \rho^{3/2}} {\bf j} \cdot
	\langle \xi \nabla \eta - \eta \nabla \xi \rangle
\nonumber
\\ 
&&
\;-\; {1 \over 4 \rho^{1/2}} 
\left( \mu - V - {g \over 2}\rho
	- {1 \over 4 \rho^2} 
	\left[ - 2 \rho \nabla^2 \rho + (\nabla \rho)^2 + {\bf j}^2 \right]
\right) \langle \xi^2 + \eta^2 \rangle.
\label{tadpole-xi}
\eqa
The last line in (\ref{tadpole-xi}) is proportional to the classical 
tadpole equation, which is just the first line of (\ref{tadpole-xi}).
The classical equation differs from zero by terms of order $\sqrt{\rho a^3}$.
Since the expectation value in the last line of (\ref{tadpole-xi}) is  
also of order $\sqrt{\rho a^3}$, the last line is
second order in $\sqrt{\rho a^3}$ and can be omitted.  
Alternatively, it can be cancelled by multiplying (\ref{tadpole-xi}) by 
$1 + \langle \xi^2 + \eta^2 \rangle/(4 \rho)$
and keeping only those terms that are first order in the expectation values.
To complete the derivation of our equations for $\rho$ and ${\bf j}$, 
we must calculate the expectation values that appear in the remaining terms of
(\ref{tadpole-xi}) and express them in terms of $\rho$ and ${\bf j}$.

\section{Thomas-Fermi expansion}
\label{sec:Thomas-Fermi}

In the previous section, the 
semiclassical tadpole equation for $\xi$ was
evaluated at a point ${\bf r}_0$ where $w$ vanishes
and expressed in terms of $\rho$, ${\bf j}$, and expectation values
of operators.
In the resulting equation (\ref{tadpole-xi}),
the operators constructed out of quantum fields $\xi$ and $\eta$ that satisfy
(\ref{xi-HF}) and (\ref{eta-HF}).  In this section,
we use the Thomas-Fermi expansion to 
evaluate the expectation values and reduce (\ref{tadpole-xi})
to a partial differential equation involving $\rho$ and ${\bf j}$ only.

The expectation values in (\ref{tadpole-xi}) are functionals of the sources 
$X$, $Y$, and $R$ given in (\ref{X-def}), (\ref{Y-def}), and (\ref{R-def}).
We would like to expand the expectation values in powers of the sources
and their gradients.  The length scale for significant variations 
in the sources is much greater than the coherence length 
$(g \rho)^{-1/2}$.
The expansion of the expectation values in powers of gradients of the sources
is possible if the expectation values receive significant contributions 
only from modes of the quantum fields 
$\xi$ and $\eta$ that have wavelengths of order 
$(g \rho)^{-1/2}$ or less.
This can be guaranteed by imposing an infrared cutoff 
that eliminates the contribution from modes with 
much longer wavelengths. 
A dependence on the infrared cutoff indicates a sensitivity 
to length scales much greater than the coherence length.
We will find that the dependence on the 
infrared cutoff cancels when (\ref{tadpole-xi})
is expressed in terms of $\rho$ and ${\bf j}$ only.

Our infrared cutoff allows an expectation value at the point ${\bf r}$
to be expanded in powers of gradients of the sources
at the point ${\bf r}$, with coefficients that are functions of 
$X$, $Y$, and $R$ at the point ${\bf r}$.  
A further expansion in powers of  $X$, $Y$, and $R$
is possible if these sources are at least first order in 
either the gradient expansion or in $\sqrt{\rho a^3}$.
This is certainly not true in general,
but we can make it true at a specific point ${\bf r}_0$
by a judicious choice of the arbitrary parameter $\Lambda$.
Note that the source $R$ in (\ref{R-def}) vanishes at the point ${\bf r}_0$
if the $U(1)$ symmetry has been used to set $w({\bf r}_0)=0$.
The sources $X$ and $Y$ in (\ref{X-def}) and (\ref{Y-def})
do not appear to be higher order in the gradient expansion or 
in $\sqrt{\rho a^3}$.
However, since the expectation values are 
already of order $\sqrt{\rho a^3}$
and we are keeping only terms to first order 
in $\sqrt{\rho a^3}$,
we can use the classical equation $T=0$ to simplify these sources:
\bqa
X &=& \Lambda^2 - g v^2 - {1 \over v} \nabla^2 v,
\label{X-simple}
\\
Y &=& - g w^2 - {1 \over v} \nabla^2 v.
\label{Y-simple}
\eqa
  From (\ref{Y-simple}), we see that $Y({\bf r}_0)$ is second order in 
the gradient expansion, since $w$ vanishes at the point ${\bf r}_0$.  
We can arrange that $X({\bf r}_0)$ also be second order in 
the gradient expansion by a judicious choice of $\Lambda$.
A convenient choice is  
\beq
\Lambda^2 \;=\; g \rho({\bf r}_0).
\label{Lambda}
\eeq
Any other choice that differs from (\ref{Lambda}) by terms 
that are higher order in the gradient expansion or in $\sqrt{\rho a^3}$
will lead to the same final equation 
for $\rho$ and ${\bf j}$.

Having imposed an infrared cutoff and chosen a specific value for $\Lambda$,
expectation values can be expanded in powers of 
$X$, $Y$, $R$, and their derivatives in a neighborhood of the point 
${\bf r}_0$:
\bqa
\langle \xi^2 \rangle &=& 
a_0 \;+\; a_1 X \;+\; a_2 Y \;+\; a_3 \nabla^2 X \;+\; a_6 \nabla^2 Y
\nonumber
\\
&& \;+\; a_4 X^2 \;+\; a_5 (\nabla X)^2 
\;+\; a_7 R^2 \;+\; a_8 (\nabla R)^2 + ...,
\label{xixi}
\\ 
\langle \eta^2 \rangle &=&
b_0 \;+\; b_1 X \;+\; b_2 Y \;+\; b_3 \nabla^2 X \;+\; b_6 \nabla^2 Y  
\nonumber
\\
&& \;+\; b_4 X^2 \;+\; b_5 (\nabla X)^2 
\;+\; b_7 R^2 \;+\; b_8 (\nabla R)^2 + ...,
\label{etaeta}
\\
\langle \xi \eta \rangle &=& 
c_1 \nabla^2 R \;+\; c_2 \nabla X \cdot \nabla R \;+\; ...,
\label{xieta}
\\
\langle \xi \nabla \eta - \eta \nabla \xi \rangle &=&
d_1 \nabla R \;+\;  ... .
\label{xigradeta}
\eqa
The coefficients $a_i$, $b_i$, $c_i$, and $d_i$ 
are functions of $\Lambda$. They
can be obtained by calculating Feynman diagrams 
using the methods illustrated in Appendix~\ref{app:Diagrams}.
These calculations are streamlined by using dimensional 
regularization to control the infrared and ultraviolet divergences 
that appear in the individual diagrams.  Analytic expressions 
for the dimensionally regularized integrals are given in 
Appendix~\ref{app:Integrals}.  The final results for the coefficients
are given in Appendix~\ref{app:Coefficients}.
We have explicitly shown only those terms in 
(\ref{xixi})--(\ref{xigradeta}) that are required to calculate
the condensate equation, the number density, and the current density
through second order in the gradient expansion.
Because of time reversal symmetry, 
$\langle \xi^2 \rangle$ and $\langle \eta^2 \rangle$
contain only terms that are even in $R$, while $\langle \xi \eta \rangle$
and $\langle \xi \nabla \eta - \eta \nabla \xi \rangle$ 
contain only terms that are odd in $R$.
Differentiating the sum of (\ref{xixi}) and (\ref{etaeta}) 
and keeping only those terms that contribute through second order 
in the gradient expansion at the point ${\bf r}_0$, we obtain
\bqa
\nabla \langle \xi^2 + \eta^2 \rangle &=& 
(a_1 + b_1) \nabla X,
\\ 
\nabla^2 \langle \xi^2 + \eta^2 \rangle &=&
(a_1 + b_1) \nabla^2X \;+\; (a_2 + b_2) \nabla^2 Y
\nonumber
\\
&& \;+\; 2 (a_4 + b_4) (\nabla X)^2 \;+\; 2(a_7 + b_7) (\nabla R)^2. 
\label{nabla2xixi}
\eqa

We proceed to express $X$, $Y$, and $R$ and their derivatives 
at the point ${\bf r}_0$ in terms of $\rho$ and ${\bf j}$.
After differentiating (\ref{X-simple}), (\ref{Y-simple}), and (\ref{R-def})
and evaluating them at ${\bf r}_0$,
we can use (\ref{list1})--(\ref{list6}) to eliminate
$v$, $w$, and their derivatives.  Since $X$, $Y$, and $R$ appear only
in expectation values that are already of order $\sqrt{\rho a^3}$,
the equations (\ref{list1})--(\ref{list6})
can be simplified by dropping the expectation values:
\bqa
v({\bf r}_0) &=& \rho^{1/2},
\\
\nabla v({\bf r}_0) &=& {1 \over 2 \rho^{1/2}} \nabla \rho,
\\
\nabla^2 v({\bf r}_0) &=& - {1 \over 4 \rho^{3/2}} 
\left[ - 2 \rho \nabla^2 \rho + (\nabla \rho)^2 + {\bf j}^2 \right],
\\
w({\bf r}_0) &=& 0, 
\\
\nabla w({\bf r}_0) &=& 
{1 \over 2 \rho^{1/2}} {\bf j},
\\
\nabla^2 w({\bf r}_0) &=& 
{1 \over 2 \rho^{1/2}} \nabla \cdot {\bf j}.
\eqa
The resulting expressions for the 
sources $X$, $Y$ and $R$ and their derivatives are
\bqa
X({\bf r}_0) &=& {1 \over 4 \rho^2} 
\left[ - 2 \rho \nabla^2 \rho + (\nabla \rho)^2 + {\bf j}^2 \right],
\label{X-0}
\\
\nabla X({\bf r}_0) &=& -g  \,\nabla \rho,
\\
\nabla^2 X({\bf r}_0)&=&
{g \over 2 \rho} \left[ - 2 \rho \nabla^2 \rho + {\bf j}^2 \right],
\\
Y({\bf r}_0) &=& {1 \over 4 \rho^2} 
\left[- 2 \rho \nabla^2 \rho + (\nabla \rho)^2 +  {\bf j}^2 \right],
\\
\nabla Y({\bf r}_0) &=& 0,
\\
\nabla^2 Y({\bf r}_0) &=& - {g \over 2 \rho} {\bf j}^2,
\\
R({\bf r}_0) &=& 0,
\\
\nabla R({\bf r}_0)&=&-\frac{g}{2}\,{\bf j},
\label{gradR-0}
\\
\nabla^2R({\bf r}_0) &=&
- {g \over 2 \rho} \left[ \nabla \rho \cdot {\bf j}
			+ \rho \nabla \cdot {\bf j} \right].
\label{grad2R-0}
\eqa
Inserting the expressions (\ref{xixi})--(\ref{nabla2xixi}) 
for the expectation values into (\ref{tadpole-xi}),
using the expressions (\ref{X-0})--(\ref{grad2R-0}) for the sources 
and their derivatives,
and truncating at second order in the gradient expansion, the semiclassical
tadpole equation for $\xi$ at the point ${\bf r}_0$ reduces to
\bqa
0 &=& 
\left( \mu - V - {g \over 2} \rho 
	- {1 \over 4 \rho^2} 
	\left[ -2 \rho \nabla^2 \rho + (\nabla \rho)^2 + {\bf j}^2 \right]
\right)	\rho^{1/2}
\nonumber
\\ 
&& \;-\; {a_0 \over 2} g \rho^{1/2}
\;+\; \left[ (a_0+b_0) + (2a_1+a_2+b_1) \Lambda^2
		+ 2a_3 \Lambda^4 \right] 
	{1 \over 4 \rho^{3/2}} \nabla^2 \rho
\nonumber
\\ 
&& \;-\; \left[ 2(a_0+b_0) + (3a_1+a_2+2b_1) \Lambda^2
	+4(a_4+b_4) \Lambda^4 + 4a_5 \Lambda^6 \right]
	{1 \over 8 \rho^{5/2}} (\nabla \rho)^2
\nonumber
\\ 
&&
\;-\; \left[ 2(a_0+b_0) + (2a_1+b_1-b_2+2d_1) \Lambda^2 
\right.
\nonumber
\\
&& \left. \hspace{1cm}
	+ (2a_3-2a_6+a_7+b_7) \Lambda^4 + a_8 \Lambda^6 \right]
	{1 \over 8 \rho^{5/2}} {\bf j}^2.
\label{tadpole-coeff}
\eqa
We have omitted the last term in (\ref{tadpole-xi}), because it is 
second order in $\sqrt{\rho a^3}$.

Inserting the coefficients $a_i$, $b_i$, and $d_1$ from 
Appendix~\ref{app:Coefficients}
into (\ref{tadpole-coeff}), using the expression for $\Lambda^2$ in 
(\ref{Lambda}), and multiplying by $-\rho^{-1/2}$,
the semclassical tadpole equation reduces to 
\bqa
\mu &=&
V \;+\; {g \over 2} \rho 
\;+\; {1 \over 4 \rho^2} 
	\left[ - 2 \rho \nabla^2 \rho +  (\nabla \rho)^2 + {\bf j}^2 \right]
\nonumber
\\
&& \;+\; {1 \over 12 \pi^2} (g^3 \rho)^{1/2}
\left( g \rho - {17 \over 192 \rho^2} 
	\left[ - 4 \rho \nabla^2 \rho + (\nabla \rho)^2 \right] \right).
\label{final1}
\eqa
This is an algebraic relation between the values of $\rho$, ${\bf j}$,
and derivatives of $\rho$.  It was derived at a specific point ${\bf r}_0$
where the condensate $\phi$ is real-valued.
However a $U(1)$ symmetry transformation can be used to make 
$\phi$ real-valued at any given point ${\bf r}$,
and $\rho$ and ${\bf j}$ are independent of that phase.
Therefore (\ref{final1}) must be valid 
at any point where the semiclassical and Thomas-Fermi approximations 
can be justified.
Using dimensional analysis to reinstate the appropriate factors of 
$\hbar$ and $2 m$ and using (\ref{g-a}) to express $g$ in terms of 
the scattering length, we obtain our final result (\ref{final-rho}).

The coefficients $a_0$ and $b_0$ in (\ref{tadpole-coeff})
are cubicly ultraviolet divergent, while $a_1$, $a_2$, $b_1$, $b_2$, 
and $d_1$ are linearly ultraviolet divergent.  These power ultraviolet 
divergences are set to zero by dimensional regularization.
If we had used a momentum cutoff $\Lambda_{UV}$, 
the ultraviolet divergences would have to be cancelled 
explicitly by counterterms.  If those counterterms were chosen to be purely
cubic or linear in $\Lambda_{UV}$, then the coefficients after 
renormalization would be identical to those obtained directly
using dimensional regularization.  Thus dimensional regularization
systematicly throws away power divergences that 
would ultimately be cancelled by counterterms.

The coefficients $a_6$, $a_8$, $b_2$ and $b_7$ 
in the quantum correction proportional to ${\bf j}^2$
are logarithmicly  infrared divergent, 
and must be calculated using an infrared cutoff.  
With dimensional regularization, these logarithmic divergences appear as 
poles in $D-3$, where $D$ is the number of spatial dimensions.
However these divergences cancel in the combinations of coefficients 
that appear in (\ref{tadpole-coeff}).  The cancellation 
of the infrared divergences indicates that our final equation 
(\ref{final1}) is 
insensitive to length scales much greater than the coherence length.  
To obtain the cancellation of infrared divergences,
it was crucial to have omitted the last term in (\ref{tadpole-xi}).
That term is infrared divergent, but the 
classical equations can be used to show that it is second order in 
$\sqrt{\rho a^3}$.  The infrared divergence can be cancelled only 
by other terms that are second order in $\sqrt{\rho a^3}$.
All such terms have been systematically dropped.

There is a remarkable cancellation of the correction 
proportional to $(g^3 \rho)^{1/2} {\bf j}^2/\rho^2$ in (\ref{tadpole-coeff}).
Because of this cancellation, the current ${\bf j}$ does not 
appear at all in the semiclassical correction term in (\ref{final1}).
We have no explanation for this cancellation.

The semiclassical correction term in (\ref{final1}) is a functional of
$\rho$ and ${\bf j}$ that includes a multiplicative factor of 
$(g^3 \rho)^{1/2}$.  The Thomas-Fermi approximation has been used to 
expand the cofactor of $(g^3 \rho)^{1/2}$ to second order in $\xi/R$,
where $\xi = (g \rho)^{-1/2}$ is the coherence length 
and $R$ is the length scale for significant variations in $\rho$.  
Since $\xi$ scales like $\rho^{-1/2}$, the Thomas-Fermi expansion 
breaks down in regions where $\rho$ approaches zero,
such as near the edge of the condensate or near the core of a vortex.
In such a region, the cofactor of $(g^3 \rho)^{1/2}$ in (\ref{final1}) 
becomes a nonlocal functional that depends on the values of 
$\rho$ and ${\bf j}$ at points within a distance of order $R$.
However the multiplicative factor of $(g^3 \rho)^{1/2}$
makes the semiclassical correction negligible in regions where $\rho \to 0$.
If $(g^3 \rho)^{1/2}/(12 \pi^2)$ is sufficiently small, then
each term in the semiclassical correction in (\ref{final1})
will be small compared to one of the terms in the classical equation.
It therefore does no harm to include these terms.
Thus the equation (\ref{final1}) can be used everywhere, 
even near the edge of the condensate or near the core of a vortex.

The chemical potential $\mu$ is the change in the total energy if a 
single atom is added to the condensate.  The terms on the right side
of (\ref{final1}) therefore have simple physical interpretations 
as contributions to the energy of that additional atom.  
In the mean-field approximation, 
that energy consists of the potential energy $V$, the interaction energy
$g \rho/2$, and the gradient energy.  
The signs of the semiclassical corrections 
are such as to increase the interaction energy 
and decrease the gradient energy.

\section{Condensate equation}
\label{sec:Condensate}

In this section, we calculate the expectation values in the 
semiclassical equation (\ref{SC-phi})
for the condensate to second order in the gradients of $\phi$
using an infrared cutoff.  The condensate equation 
is expressed in the form of a partial differential equation for $\phi$
that depends logarithmicly on the infrared cutoff.
We will find that if $\phi$ is eliminated in favor of $\rho$ and ${\bf j}$,
the dependence on the infrared cutoff cancels and we recover the
equation (\ref{final1}).

The expectation values appearing in the condensate equation
(\ref{SC-phi}) are 
expectation values of operators involving 
real quantum fields $\xi$ and $\eta$ that satisfy
(\ref{xi-HF}) and (\ref{eta-HF}):
\bqa
\langle \tilde\psi^\dagger \tilde\psi \rangle
&=& {1 \over 2} \langle \xi^2 + \eta^2 \rangle,
\label{normal-HF}
\\
\langle \tilde\psi \tilde\psi \rangle
&=& {1 \over 2} \langle \xi^2 - \eta^2 \rangle
\;+\; i  \langle \xi \eta \rangle.
\label{anom-HF}
\eqa
Expansions for $\langle \xi^2 \rangle$, $\langle \eta^2 \rangle$,
and $\langle \xi \eta \rangle$
in powers of $X$, $Y$, $R$, and their derivatives are given in 
(\ref{xixi})--(\ref{xieta}). 
These expansions are valid in a neighborhood of a point ${\bf r}_0$
where $w$ vanishes, provided that we make a suitable choice for the
arbitrary parameter $\Lambda$.  If we wish to express the expectation values
(\ref{normal-HF}) and (\ref{anom-HF}) in terms of the condensate,
the simplest choice is 
\beq
\Lambda^2 \;=\; g v^2({\bf r}_0).
\label{Lambda-v}
\eeq
This differs from the choice (\ref{Lambda}) made in the previous section 
by terms of order $\sqrt{\rho a^3}$.  The choice (\ref{Lambda-v})
leads to the same final result 
if we truncate at the same orders in $\sqrt{\rho a^3}$ and in $\xi/R$.
After differentiating the expressions (\ref{X-simple}),
(\ref{Y-simple}), and (\ref{R-def}) for the sources and 
evaluating them at the point ${\bf r}_0$, they can be simplified using
$w({\bf r}_0)=0$.  Inserting the 
resulting expressions into
(\ref{xixi})--(\ref{xigradeta}), we obtain
\bqa
\langle \xi^2\rangle &=& a_0 
\;-\; \left[ a_1 + a_2 + 2 a_3 \Lambda^2 \right] \nabla^2 v/v
\nonumber
\\ 
&& \;-\; 2 \left[ a_3 - 2 a_5 \Lambda^2 \right] g (\nabla v)^2
\;-\; \left [ 2 a_6 - a_8 \Lambda^2 \right] g (\nabla w)^2,
\label{mat1}
\\
\langle \eta^2 \rangle &=& b_0
\;-\; \left[ b_1 + b_2 + 2 b_3 \Lambda^2 \right] \nabla^2 v/v
\nonumber
\\ 
&& \;-\; 2 \left[ b_3 - 2 b_5 \Lambda^2 \right] g (\nabla v)^2 
\;-\; \left [ 2 b_6 - b_8 \Lambda^2 \right] g (\nabla w)^2,
\label{mat2}
\\
\langle \xi \eta \rangle &=& - c_1 g v \nabla^2 w
\;-\; 2 \left[c_1 - c_2 \Lambda^2 \right] g \nabla v \cdot \nabla w,
\label{mat3}
\\
\langle \xi \nabla \eta - \eta \nabla \xi \rangle &=& 
- d_1 g v \nabla w.
\label{mat4}
\eqa

Some of the coefficients in the expansions (\ref{mat1}) and (\ref{mat2})
are infrared divergent and must be calculated using an infrared cutoff.
The coefficients $a_6$ and $a_8$ are logarithmicly infrared divergent,
but the divergence cancels in the combination $2 a_6 - \Lambda^2 a_8$ 
that appears in (\ref{mat1}).
The coefficients $b_6$ and $b_8$ are quadraticly infrared divergent,
but the leading divergences cancel in the combination $2 b_6 - \Lambda^2 b_8$ 
that appears in (\ref{mat2}), leaving a logarithmic divergence.
There are also logarithmic infrared divergences in 
$b_2$, $b_3$, $b_5$, and $b_7$.
We choose to use dimensional regularization for the infrared cutoff.
Logarithmic infrared divergences 
appear as poles in $D-3$, where $D$ is the number of dimensions.
The expressions for the infrared divergent integrals can be simplified by 
trading the poles in $D-3$ for logarithms
of a momentum scale $\mu_{IR}$ defined by
\beq
\log {\Lambda^2 \over \mu^2_{IR}} \;=\; 
{2 \over D-3} + \log {4 \Lambda^2 \over \pi \mu^2} + \gamma,
\label{pole-log}
\eeq
where $\gamma$ is Euler's constant and $\mu$ is the momentum scale 
introduced by dimensional regularization.
The momentum scale $\mu_{IR}$ can be interpreted as a conventional
infrared momentum cutoff.
Using the values for the coefficients given in Appendix~\ref{app:Coefficients}
and using the choice (\ref{Lambda-v}) for $\Lambda$,
the expectation values (\ref{mat1})--(\ref{mat3}) reduce to
\bqa
\langle \xi^2  \rangle &=& {1 \over 6 \pi^2} g^{3/2} v^3
\;-\; {7 \over 288 \pi^2} g^{1/2} (\nabla v)^2/v
\nonumber
\\ 
&& \;+\; {11 \over 144 \pi^2} 
	g^{1/2} \left( \nabla^2 v + (\nabla w)^2/v \right) ,
\label{xixi-0}
\\
\langle \eta^2 \rangle &=& - {1 \over 12 \pi^2} g^{3/2} v^3
\;-\; {1 \over 64 \pi^2}
	\left[ \log{g v^2 \over \mu_{IR}^2} - {62 \over 9} \right]
	g^{1/2} (\nabla v)^2/v 
\nonumber
\\ 
&& \;-\; {5 \over 96 \pi^2}
	\left[ \log{g v^2 \over \mu_{IR}^2} - {44 \over 15} \right]
	 g^{1/2} \left( \nabla^2 v + (\nabla w)^2/v \right),
\label{etaeta-0}
\\
\langle \xi \eta \rangle &=& 
- {1 \over 12 \pi^2} g^{1/2} 
	\left( \nabla^2 w + \nabla v \cdot \nabla w/v \right),
\label{xieta-0}
\\
\langle \xi \nabla \eta - \eta \nabla \xi \rangle &=& 
{1 \over 12 \pi^2} g^{3/2} v^2 \nabla w.
\label{xigradeta-0}
\eqa
The infrared--divergent coefficients in (\ref{etaeta-0}) 
indicate that $\langle \eta^2 \rangle$ is logarithmicly 
sensitive to length scales much greater than the coherence length.

Inserting  (\ref{xixi-0})--(\ref{xieta-0}) into 
(\ref{normal-HF}) and ({\ref{anom-HF}),
we obtain expressions for the noncondensate density
and the anomalous density that
hold at a particular point ${\bf r}_0$ where $w({\bf r}_0)$ vanishes.
We can use the $U(1)$ symmetry 
to deduce general expressions for these expectation values
in terms of $\phi$ and gradients of $\phi$.
The noncondensate density (\ref{normal-HF}) 
is invariant under the $U(1)$ symmetry.
At leading order in the gradient expansion, it must be a function of
$|\phi|^2 = \phi^*\phi$, which reduces to $v^2$ at the point ${\bf r}_0$.
Thus the $U(1)$--invariant expression for the $v^3$ term in  
(\ref{normal-HF}) is simply $|\phi|^3$. 
At second order in the gradient expansion, the noncondensate density
can be expressed as a linear combination of 
$\nabla \phi^* \cdot \nabla \phi$, 
$\nabla^2(\phi^* \phi)$, and 
$\left[\nabla (\phi^* \phi) \right]^2$,
with coefficients that are functions of $|\phi|$.
Any $U(1)$--invariant expression that is second order in gradients
of $\phi$
can be expressed as a linear combination of these three terms.
We can deduce the coefficients of those terms in
$\langle \tilde\psi^\dagger \tilde\psi \rangle$ by evaluating the
linear combination at the 
point ${\bf r}_0$ and comparing to (\ref{normal-HF}).
The resulting $U(1)$--invariant expression for the noncondensate density is
\bqa
\langle \tilde\psi^\dagger \tilde\psi \rangle
&=& {1 \over 24 \pi^2} g^{3/2} |\phi|^3
\;-\; {5 \over 384 \pi^2}
	\left[ \log{g |\phi|^2 \over \mu_{IR}^2} - {22 \over 5} \right] 
	g^{1/2} |\phi|^{-1} \nabla^2(\phi^* \phi)
\nonumber
\\
&& \;+\; {7 \over 1536 \pi^2}
	\left[ \log{g |\phi|^2 \over \mu_{IR}^2} - 4 \right] 
	g^{1/2} |\phi|^{-3} \left[\nabla (\phi^* \phi) \right]^2.
\label{normal-U1}
\eqa
The anomalous density (\ref{anom-HF}) is a functional of $\phi$ 
that transforms like $\psi^2$ under the $U(1)$ symmetry (\ref{U1-sym}).
At leading order in the gradient expansion, it must have the form
$\phi^2$ multiplied by a function of $|\phi|$.
Thus the $U(1)$--covariant expression for the $v^3$ term in 
(\ref{anom-HF}) is $|\phi| \phi^2$.  
At second order in the gradient expansion, there are five
independent terms that transform like $\psi^2$ under the $U(1)$ symmetry:
$\phi \nabla^2 \phi$,
$(\nabla \phi)^2$,
$\phi^2 \nabla \phi^* \cdot \nabla \phi$, 
$\phi^2 \nabla^2 (\phi^* \phi)$,  and 
$\phi^2 \left[\nabla (\phi^* \phi) \right]^2$.
The anomalous density (\ref{anom-HF}) must be a linear 
combination of these terms with coefficients that are functions of $|\phi|$.
We can determine the coefficients by evaluating this linear combination 
at the point ${\bf r}_0$ and matching with 
(\ref{anom-HF}).  The resulting $U(1)$--covariant expression 
for the anomalous density is
\bqa
\langle \tilde\psi \tilde\psi \rangle
&=& {1 \over 8 \pi^2} g^{3/2} |\phi| \phi^2
\;-\; {1 \over 24 \pi^2} g^{1/2} |\phi|^{-3} 
	\left[ 2 |\phi|^2 \phi \nabla^2 \phi
		+ |\phi|^2 (\nabla \phi)^2
		+ 3 \phi^2 \nabla \phi^* \cdot \nabla \phi \right]
\nonumber
\\
&& \;+\; {5 \over 384\pi^2}
	\left[ \log{g |\phi|^2 \over \mu_{IR}^2} + {26 \over 15} \right] 
	g^{1/2} |\phi|^{-3} \phi^2 \nabla^2(\phi^*\phi)
\nonumber
\\ 
&& \;-\; {7 \over 1536\pi^2}
       \left[ \log{g |\phi|^2 \over \mu_{IR}^2} - {64 \over 21} \right] 
	g^{1/2} |\phi|^{-5} \phi^2 \left[\nabla (\phi^* \phi) \right]^2.
\label{anom-U1}
\eqa

Inserting the expectation values (\ref{normal-U1}) and (\ref{anom-U1})
into (\ref{SC-phi}), we obtain the semiclassical equation for the condensate 
to second order in the Thomas-Fermi expansion:
\bqa
0 &=& (\mu - V) \phi
\;-\; {g\over 2} \left[ 1 + {5 \over 24 \pi^2} g^{3/2} |\phi| \right] 
	|\phi|^2 \phi
\;+\; \left[ 1 + {1 \over 24 \pi^2} g^{3/2} |\phi| \right] \nabla^2 \phi 
\nonumber
\\
&& \;+\; {1 \over 48 \pi^2} g^{3/2} |\phi|^{-1}
	\left[ \phi^* (\nabla \phi)^2 
		+ 3 \phi \nabla \phi^* \cdot \nabla \phi \right]
\nonumber
\\
&& \;+\; {5 \over 768 \pi^2} 
	\left[ \log{g |\phi|^2 \over \mu_{IR}^2} - {158 \over 15} \right]
	g^{3/2} |\phi|^{-1} \phi \nabla^2 (\phi^* \phi)
\nonumber
\\
&& \;-\; {7 \over 3072 \pi^2} 
	\left[ \log {g |\phi|^2 \over \mu_{IR}^2} - {104 \over 21} \right]
	g^{3/2} |\phi|^{-3} \phi \left[ \nabla (\phi^* \phi) \right]^2.
\label{phi-U1}
\eqa
The infrared--divergent coefficients in the last two terms of (\ref{phi-U1})  
indicate that, at second order in the gradient expansion,
the condensate is logarithmicly 
sensitive to length scales much greater than the coherence length.

We can also obtain $U(1)$--invariant expressions for the  
number density and the current density in terms of the condensate $\phi$.
For the number density, this is simply a matter of inserting the 
noncondensate density (\ref{normal-U1}) into (\ref{rho-HF}):
\bqa
\rho &=& 
\left[ 1 + {1 \over 24 \pi^2} g^{3/2} |\phi| \right] |\phi|^2 
\;-\; {5 \over 384 \pi^2}
	\left[ \log{g |\phi|^2 \over \mu_{IR}^2} - {22 \over 5} \right] 
	g^{1/2} |\phi|^{-1} \nabla^2(\phi^* \phi)
\nonumber
\\
&& \;+\; {7 \over 1536 \pi^2}
	\left[ \log{g |\phi|^2 \over \mu_{IR}^2} - 4 \right] 
	g^{1/2} |\phi|^{-3} \left[\nabla (\phi^* \phi) \right]^2.
\label{rho-U1}
\eqa
To obtain the current density, we need 
the $U(1)$--invariant expression for the expectation value 
(\ref{xigradeta-0}):
\beq
\langle \xi \nabla \eta - \eta \nabla \xi \rangle \;=\; 
- {i \over 24 \pi^2} g^{3/2} |\phi|
	\left( \phi^* \nabla \phi -  \phi \nabla \phi^* \right).
\label{xigradeta-U1}
\eeq
Inserting this into (\ref{j-HF}), we find that the current reduces to
\beq
{\bf j} \;=\;
-i \left[ 1 \;+\; {1 \over 24 \pi^2} g^{3/2} |\phi| \right]
	\left( \phi^* \nabla \phi - \phi \nabla \phi^* \right).
\label{j-U1}
\eeq

Given the condensate equation (\ref{phi-U1}) and the expressions 
for $\rho$ and ${\bf j}$ in (\ref{rho-U1}) and (\ref{j-U1}),
it is straightforward to derive the partial differential equations for 
$\rho$ and ${\bf j}$ given in (\ref{j-cont}) and (\ref{final1}).
It is convenient to first multiply (\ref{phi-U1}) by $\phi^*$.
The imaginary part of the resulting equation
immediately gives the continuity equation $\nabla \cdot {\bf j} = 0$,
where ${\bf j}$ is given in (\ref{j-U1}).
In the real part of the equation,
we can use (\ref{rho-U1}) and (\ref{j-U1}) to eliminate $\phi$ in favor of
$\rho$ and ${\bf j}$.
We find that all the infrared divergences
can be organized into a term that is proportional to the classical 
equation for $\rho$ and ${\bf j}$.  That term is
second order in $\sqrt{\rho a^3}$ and can be consistently deleted.
The remaining terms reduce to (\ref{final1}) up to an overall 
multiplicative factor.  The fact that 
the $U(1)$--invariant form of the condensate equation (\ref{phi-U1})
reproduces the differential equation (\ref{final1}) for $\rho$ and ${\bf j}$
provides a stringent check on both results.

\section{Conclusions}
\label{sec:Con}

In this paper, we have calculated the corrections from 
quantum field fluctuations to the mean-field equations for a 
time-independent state of a Bose-Einstein condensate.
We used only controlled approximations that correspond to 
truncations of systematic expansions in small quantities:
the semiclassical approximation, which is the expansion to 
first order in $\sqrt{\rho a^3}$, and the Thomas-Fermi approximation,
which is an expansion in powers of $\xi/R$.
The consistency of these approximations is guaranteed by the fact 
that they are controlled by small expansion parameters.

By integrating out the quantum field fluctuations,
we obtained remarkably simple self-contained equations for the 
condensate $\phi$ and for the hydrodynamic fields $\rho$ and ${\bf j}$.
The effects of quantum field fluctuations were taken into account 
through local correction terms to the partial differential equations 
of the mean-field approximation.  In the Thomas-Fermi limit,
the correction to the Gross-Pitaevskii equation is
the $|\phi|^3 \phi$ term in (\ref{final-phi}).  At second order in the 
Thomas-Fermi expansion, the  corrections to the Gross-Pitaevskii equation
can no longer be expressed in a local form.  However the effects 
of quantum field fluctuations can be taken into account by adding
local correction terms to the hydrodynamic form of the 
mean-field equations.  The resulting equations are the continuity equation
(\ref{j-cont}) and the nontrivial equation (\ref{final-rho}).
The numerical solution of these partial differential equations is much 
simpler than solving the equations of the Hartree-Fock-Bogoliubov 
approximation.

The Thomas-Fermi approximation has also been used in a recently
developed variational Thomas-Fermi theory of a nonuniform
Bose-Einstein condensate \cite{T-T-H}.  In the case of atoms that
interact only through the S-wave scattering length, this approach
leads to results similar to ours in the Thomas-Fermi limit.
The authors were unable to proceed beyond leading order
in the Thomas-Fermi expansion because of the breakdown 
of the gradient expansion.  We avoided this problem by changing to the
hydrodynamic variables $\rho$ and ${\bf j}$.

Our methods can be readily extended to 
time-dependent states of a Bose-Einstein condensate.
The resulting equations would describe the collective excitations
of a Bose-Einstein condensate, including the leading effects of quantum 
field fluctuations.  A particularly simple 
application of these equations would be to study the frequencies 
for small oscillations of the condensate.  In 
the mean-field approximation and in the Thomas-Fermi limit,
the oscillation frequencies for a
spherically-symmetric harmonic oscillator potential $V(r) = m \omega^2r^2/2$ 
are particularly simple \cite{Stringari}.
The  frequencies are independent of the scattering length $a$
and ratios of frequencies are independent of $\omega$.
Quantum field fluctuations give 
fractional corrections to the frequencies proportional to 
$N^{1/5}(a^2 m \omega/\hbar)^{3/5}$.  
It would be very interesting to calculate
these changes in the frequencies and see if they can be measured in
experiments. 

Up to this point, we have only applied our methods to the 
relatively simple problem of a Bose-Einstein condensate at
zero temperature.  
It may be possible to generalize our methods to the case of 
nonzero temperatures $T$ that are below the critical temperature 
$T_c$ for the Bose-Einstein phase transition.
The temperature introduces a new length scale into the problem, 
the thermal wavelength $\hbar/\sqrt{2mkT}$, and therefore a new
dimensionless ratio $mkT/(\hbar^2 \rho_0 a)$.
The condensate number density $\rho_0$ decreases to 0 as $T$ approaches $T_c$.
Since the coherence length $\xi$ scales like $(\rho_0 a)^{-1/2}$, 
the Thomas-Fermi expansion
in $\xi/R$ breaks down for large enough $T$.  However the product of
$\sqrt{\rho_0 a^3}$ and $mkT/(\hbar^2 \rho_0 a)$ remains small at all
temperatures below the critical region near $T_c$ \cite{F-S}. 
This quantity can therefore be used as an expansion parameter 
for a controlled approximation.
One may be able to use this controlled approximation to 
systematically integrate out the effects of short-wavelength fluctuations 
around the mean field.  It would be interesting to see if this can 
lead to a simplification of the equations for nonzero temperature 
that is comparable to the simplification that we have achieved at 
zero temperature.

\subsection*{Acknowledgments}

This work was supported in part by the U.~S. Department of Energy,
Division of High Energy Physics, under Grant DE-FG02-91-ER40690
and by a Faculty Development Grant from the Physics Department of The 
Ohio State University.
J.O.A. was also supported in part 
by a Fellowship 
from the Norwegian Research Council (project 124282/410).

\appendix
\renewcommand{\theequation}{\thesection\arabic{equation}}

\section{Integrals}
\label{app:Integrals}
\setcounter{equation}{0}

In this Appendix, we give analytic expressions for the energy integrals 
and the momentum integrals that are required to calculate the
coefficients in the expansions of the expectation values 
(\ref{xixi})--(\ref{xigradeta}) in powers of the sources 
and their derivatives.  
We use dimensional regularization to cut off any infrared or
ultraviolet divergences in the momentum integrals.

The energy integrals 
have poles at $\omega = \pm \epsilon$, where 
$\epsilon = k \sqrt{k^2 + \Lambda^2}$.  The poles come from Feynman 
propagators, which are defined by an $i 0^+$ prescription.
The integrals can be evaluated using contour integration.
The specific integrals that are required are
\begin{eqnarray}
\int{d \omega \over 2 \pi} 
	{1 \over (\omega^2 - \epsilon^2 + i 0^+)^n}
&=& i (-1)^{n+1} {(-\mbox{$1\over2$})_n \over (n-1)!} 
	{1 \over \epsilon^{2n-1}} \,,
\label{int-omega}
\\ 
\int{d \omega \over 2 \pi} 
	{\omega^2 \over (\omega^2 - \epsilon^2 + i 0^+)^{n+1}}
&=& i (-1)^{n+1} {(-\mbox{$1\over2$})_n \over 2 \; n!} 
	{1 \over \epsilon^{2n-1}} \,.
\label{int-omega'}
\end{eqnarray}
where $(z)_n$ is the Pochhammer symbol:  $(z)_n = \Gamma(z+n)/\Gamma(z)$.
 
Some of the momentum integrals are ultraviolet divergent or
infrared divergent or both.  We choose to use
dimensional regularization to regularize both 
ultraviolet and infrared divergences. 
This involves calculating the integral as an analytic function 
of the number of dimensions $D$ and analytically continuing it to $D=3$.
The analytic continuation sets power ultraviolet divergences
and power infrared divergences equal to 0, 
but logarithmic divergences appear as poles in $D-3$.
It is convenient to take the integration measure to be
$\mu^{3-D}\int d^Dk/(2 \pi)^D$,
where $\mu$ is an arbitrary momentum scale.
The prefactor $\mu^{3-D}$ has been inserted 
so that the regularized integral has the same 
engineering dimensions as in $D=3$.

The insertion of a source into a loop diagram shifts the 
loop momentum ${\bf k}$ by the momentum flowing through the source.
The gradient expansion corresponds to expanding the loop integral 
in powers of the momenta flowing through the sources.  
This expansion generates integrals over the momentum ${\bf k}$ 
with vector indices.  These integrals can be reduced to scalar
integrals by averaging over angles in $D$ dimensions
using the formulas
\bqa
\int {d^Dk \over (2\pi)^D} f(k^2) k_i &=& 0,
\label{angular}
\\
\int {d^Dk \over (2\pi)^D} f(k^2) k_i k_j &=&
{1 \over D} \delta_{ij} \int {d^Dk \over (2\pi)^D} f(k^2) k^2.
\label{angular'}
\eqa
One must avoid setting $D=3$ on the right side of (\ref{angular})
if the scalar integral over ${\bf k}$ has a logarithmic divergence.

The scalar momentum integrals can be written in the form
\begin{equation}
\label{required}
I_{m,n} \;\equiv\; 
\mu^{3-D}
\int{d^Dk \over (2 \pi)^D} {(k^2)^m \over (k \sqrt{k^2 + \Lambda^2})^n} \,,
\label{Imn-def}
\end{equation}
where $m$ and $n$ are integers.
Evaluating (\ref{Imn-def}) in a region of $D$ where it is convergent 
and then expressing it as an analytical function of $D$, we obtain
\begin{equation}
I_{m,n} \;=\; 
{ \Gamma(n - m - {D \over 2}) \Gamma({D-n \over 2} + m) 
	\over (4 \pi)^{D/2} \Gamma({n \over 2}) \Gamma({D \over 2}) } \,
	\Lambda^{D+2m-2n} \,\mu^{3-D}\,.
\label{Imn-D}
\end{equation}
These integrals satisfy the identities
\bqa
I_{m-1,n-2} - I_{m+1,n} &=& \Lambda^2 I_{m,n},
\\
(D + 2 m - n) I_{m,n} &=& n I_{m+2,n+2}.
\label{IBP}
\eqa
The first is just an algebraic relation, while the second follows from
applying integration by parts to (\ref{required}). 

One of the advantages of dimensional regularization is that the formulas
(\ref{Imn-D})--(\ref{IBP}) 
hold even if the integrals are infrared or ultraviolet divergent.
For $D=3$, $I_{m,n}$ is ultraviolet divergent if
$m$ and $n$ satisfy $m-n \ge - {3 \over 2}$,
and infrared divergent if $2m-n \le -3$.
The ultraviolet divergences are power divergences
that are set to zero by dimensional regularization.
Thus, unless $2m-n \leq -3$, we can set $D=3$ in (\ref{Imn-D}).
The integral then reduces to
\begin{equation}
I_{m,n} \;=\; 
{ \Gamma(n - m - {3 \over 2}) \Gamma({3-n \over 2} + m) 
	\over 4 \pi^2 \Gamma({n \over 2}) } \,
	\Lambda^{3+2m-2n}\,, 
\qquad n < 2m+3 \,.
\label{Imn-3}
\end{equation}
If $2m-n = -3$, $I_{m,n}$ has a logarithmic infrared divergent
for $D=3$.  With dimensional regularization, this divergence  
appears as a pole in $D-3$. 
After extracting the pole from 
the factor $\Gamma({D-3 \over 2})$ in (\ref{Imn-D}) and expanding the 
remaining factors to first order in $D-3$, we obtain
\bqa
I_{-2,-1} &=& {1 \over 4 \pi^2}
\left[ {2 \over D-3} + \log {4 \Lambda^2 \over \pi \mu^2} 
	+ \gamma - 4 \right] \Lambda,
\label{I-2-1}
\\
I_{-1,1} &=& {1 \over 4 \pi^2}
\left[ {2 \over D-3} + \log {4 \Lambda^2 \over \pi \mu^2} 
	+ \gamma - 2 \right] {1 \over \Lambda},
\\
I_{0,3} &=& {1 \over 4 \pi^2}
\left[ {2 \over D-3} + \log {4 \Lambda^2 \over \pi \mu^2} 
	+ \gamma - 4 \right] {1 \over \Lambda^3}.
\eqa
If $2m-n = -5$, $I_{m,n}$ has a quadratic infrared divergence for $D=3$, 
underneath which is a logarithmic infrared divergence.
Dimensional regularization sets the quadratic divergence to zero, 
but the logarithmic divergence appears as a pole in $D-3$. 
After extracting the pole from 
the factor $\Gamma({D-5 \over 2})$ in (\ref{Imn-D}) and expanding the 
remaining factors to first order in $D-3$, we obtain
\bqa
I_{-3,-1} &=& {1 \over 8 \pi^2}
\left[ {2 \over D-3} + \log {4\Lambda^2 \over \pi \mu^2}
	+\gamma - 1 \right] {1 \over \Lambda},
\\
I_{-2,1} &=& - {1 \over 8 \pi^2}
\left[ {2 \over D-3} + \log {4 \Lambda^2 \over \pi \mu^2} 
	+ \gamma - 3 \right] {1 \over \Lambda^3}.
\label{I-21}
\eqa
The product of $D-3$ with a pole in $D-3$ gives a nonzero result
in the limit $D \to 3$.
Thus if any of the infrared divergent integrals in (\ref{I-2-1})--(\ref{I-21})
is multiplied by a function of $D$, one must take care to expand 
that function 
to first order in $D-3$ before taking the limit $D \to 3$.

\section{Feynman diagrams}
\label{app:Diagrams}
\setcounter{equation}{0}

In this Appendix, we illustrate the use of Feynman diagrams
to calculate the coefficients in (\ref{xixi})--(\ref{xigradeta}).
In these equations, the expectation values of quantum field 
operators in the presence of the sources $X$, $Y$, and $R$ are 
expanded in powers of the sources and their gradients.
As an example, we will calculate the coefficient $c_1$
in the expansion (\ref{xieta}) for $\langle \xi \eta \rangle$.

The basic ingredients of Feynman diagrams are propagators and vertices.
It is convenient to calculate the diagrams in momentum space.
The diagonal propagators for the fields 
$\xi$ and $\eta$ are represented by solid lines and dashed lines, 
respectively.  Their Feynman rules are
\begin{eqnarray}
&&{i k^2 \over  \omega^2 - \epsilon^2(k) + i 0^+},
\label{prop-xixi}
\\
&&{i \epsilon^2(k)/k^2 \over  \omega^2 - \epsilon^2(k) + i 0^+},
\label{prop-etaeta}
\end{eqnarray}
where ${\bf k}$ is the momentum and $\omega$ is the energy 
flowing through the line.  
The off-diagonal propagator for $\xi$ and $\eta$ is
represented by a line that is half solid and half dashed.  
Its Feynman rule is
\begin{equation}
{\omega \over  \omega^2 - \epsilon^2(k) + i 0^+},
\label{prop-xieta}
\end{equation}
where $\omega$ is the energy flowing from the dashed end of the line 
toward the solid end.
Each of the operators $\xi^2$, $\eta^2$, and $\xi \eta$, and 
$\xi \nabla \eta - \eta \nabla \xi$ 
creates two outgoing lines. 
The vertex factors for the operators 
$\xi^2$, $\eta^2$, and $\xi \eta$ are 2, 2, and 1, respectively. 
The vertex factor for the operator $\xi \nabla \eta - \eta \nabla \xi$ 
is ${\bf p}- 2{\bf k}$ if the momenta flowing out 
the $\xi$ and $\eta$ lines created by the operator are
${\bf k}$ and ${\bf p}- {\bf k}$, respectively.
The only other vertices needed to calculate the one-loop diagrams 
for the expectation values of these operators are for the external
sources $X$, $Y$, and $R$.  If momentum ${\bf p}$ is flowing out of the 
diagram though one of those sources, the vertex factor is $iX({\bf p})$,
$iY({\bf p})$, or $iR({\bf p})$. 
The one-loop diagrams
are integrated over the energy $\omega$ and momentum ${\bf k}$ 
running around the loop, with the measures $\int d \omega/(2 \pi)$ and 
$\mu^{3-D} \int d^Dk/(2 \pi)^D$, 
where $D$ is the number of spatial dimensions.
Finally, there is a symmetry factor of 1/2 if the diagram has a 
reflection symmetry.

We consider the diagrams in Fig.~\ref{so}, which represent the
contributions to the expectation value $\langle \xi\eta \rangle$ 
involving one insertion of the source $R$.
If the external momentum ${\bf p}$ enters through the operator
$\xi\eta$ and exits through the source $R({\bf p})$,
the expression for the first diagram in Fig.~\ref{so} is 
\bqa
\int{d\omega\over2\pi} \mu^{3-D} \int{d^Dk\over (2\pi)^D}
	{(-\omega)\over\omega^2-\epsilon^2(|{\bf k+p}|)} iR({\bf p})
	{(-\omega)\over\omega^2-\epsilon^2(|{\bf k}|)}.
\label{inte}
\eqa
We have written the Feynman rules for each of the propagators and the vertex 
in the loop in the order in which they appear as you go counterclockwise 
around the loop.  There is an implied  $+i0^+$ prescription
in the denominator of each of the propagators.

The first step in evaluating the diagram 
is to expand the integrand in powers of the external 
momentum ${\bf p}$.
The expansion of the first denominator  in (\ref{inte})
to second order in ${\bf p}$ has the form
\begin{eqnarray}
{1 \over \omega^2 - \epsilon^2(| {\bf k} + {\bf p} |)}
&=& 
{1 \over \omega^2 - \epsilon^2}
\nonumber \\
&& \;+\; \left[
\left( k^2 + {\epsilon^2 \over k^2} \right)
	{ \left(p^2 + 2 {\bf p} \cdot {\bf k}\right) }
\;+\; 4 ({\bf p} \cdot {\bf k})^2 \right]
	{1 \over (\omega^2 - \epsilon^2)^2}
\nonumber \\
&& \;+\; 
4 \left( k^2  + {\epsilon^2 \over k^2} \right)^2
({\bf p} \cdot {\bf k})^2
	{1 \over (\omega^2 - \epsilon^2)^3} \,,
\end{eqnarray}
where $\epsilon=\epsilon(k)$ on the right hand side.
We can average
over the angles of ${\bf k}$ by substituting $k_ik_j\rightarrow 
k^2\delta_{ij}/D$ and $k_i\rightarrow 0$. The term
linear in ${\bf p}\cdot{\bf k}$ drops out, and the diagram
(\ref{inte}) reduces to
\bqa
iR({\bf p})\int{d\omega\over2\pi}
\mu^{3-D} \int{d^Dk\over (2\pi)^D}
{\omega^2\over(\omega^2-\epsilon^2)^2}
\left\{1 \;+\;
\left[ {D (k^4+\epsilon^2) + 4 k^4 \over D k^2 (\omega^2-\epsilon^2)}
	+{4(k^4+\epsilon^2)^2\over D k^2 (\omega^2-\epsilon^2)^2}
\right] p^2 
\right\}.
\eqa
We can now use the formula (\ref{int-omega'})
to evaluate the integrals over $\omega$.
This reduces the diagram to an integral over ${\bf k}$:
\bqa
R({\bf p}) \mu^{3-D} \int{d^Dk\over(2\pi)^D}
\left\{{1\over4\epsilon} \;+\;
\left[- {D (k^4+\epsilon^2) + 4 k^4 \over 16 D k^2 \epsilon^3}
	+{\left(k^4+\epsilon^2\right)^2\over 8 D k^2 \epsilon^5}
\right] p^2
\right\}.
\eqa
In terms of the integrals $I_{m,n}$ defined in (\ref{Imn-def}), this reads
\bqa
{1\over4}I_{0,1}R({\bf p})
+\left[{1\over8D}I_{3,5}-{1\over16}I_{1,3}
-{D-2\over16D}I_{-1,1}\right]p^2R({\bf p}).
\label{diagram1}
\eqa

The expression for the second diagram in Fig.~\ref{so} is
\bqa
\int{d\omega\over2\pi}
\mu^{3-D} \int{d^Dk\over(2\pi)^D}{i({\bf k}+{\bf p})^2
\over \omega^2-\epsilon^2({\bf k}+{\bf p})}iR({\bf p})
{i\epsilon^2(k)/k^2\over\omega^2-\epsilon({\bf k})^2}.
\eqa
It is calculated in the same manner and the
result is
\bqa
{1\over4}I_{0,1}R({\bf p})
+\left[{5\over8D}I_{3,5}-{3D+4\over16D}I_{1,3}
+{D-2\over16D}I_{-1,1}\right]p^2R({\bf p}).
\label{diagram2}
\eqa
The sum of the diagrams (\ref{diagram1}) and (\ref{diagram2}) is
\beq
{1 \over 2}I_{0,1}R({\bf p})
\;+\; \left[ {3\over 4D} I_{3,5} - {D+1 \over 4D} I_{1,3} \right] 
	p^2 R({\bf p}).
\eeq
Using the identity (\ref{IBP}), this reduces to
\beq
{1 \over 2} I_{0,1}R({\bf p})
- {1\over2D} I_{1,3}p^2R({\bf p}).
\eeq
After Fourier transforming to coordinate space,
we can write this as
\beq
\langle \xi  \eta({\bf x}) \rangle \;=\;
{1 \over 2}I_{0,1}R({\bf x}) 
\;+\; {1\over2D}I_{1,3}\nabla^2R({\bf x}) \;+\; \ldots ,
\eeq
where $\ldots$ represents term with more gradients of $R$
or more powers of $X$, $Y$, or $R$.
Comparing with the expansion for $\langle\xi\eta\rangle$ in (\ref{xieta}),
we can read off the coefficient $c_1 = I_{1,3}/(2D)$.
We also find that there is a term $c_0 R$ in the expansion (\ref{xieta}),
with $c_0 = I_{0,1}/2$.  This term was omitted in (\ref{xieta}),
because $c_0$ does not contribute to any of the quantities calculated 
in this paper.

\section{Coefficients}
\label{app:Coefficients}
\setcounter{equation}{0}

In this Appendix, we express the coefficients that appear in 
(\ref{xixi})--(\ref{xigradeta}) in terms of scalar momentum integrals.
The coefficients can be calculated 
by evaluating Feynman diagrams, as described in Appendix~\ref{app:Diagrams}
of Ref. \cite{Braaten-Nieto:2}.  
After using (\ref{int-omega}) and (\ref{int-omega'}) 
to integrate over the energy and (\ref{angular}) and (\ref{angular'}) 
to average over angles, 
the momentum integrals can be reduced to the scalar  
integrals $I_{m,n}$ defined in Appendix~\ref{app:Integrals}.

There are twelve coefficients that are required to calculate
$\phi$ and $\rho$ for the ground state in the semiclassical approximation 
and to second order in the Thomas-Fermi expansion.  
They were calculated previously 
in Ref.~\cite{Braaten-Nieto:2} using a large momentum cutoff 
to regularize ultraviolet divergences and a small momentum cutoff 
to regularize infrared divergences.  
Since we are using dimensional regularization 
for both infrared and ultraviolet divergences,
we need the expressions for those coefficients in $D$ dimensions.
Eight of the coefficients are given by the same expressions 
as in Ref.~\cite{Braaten-Nieto:2}:
\bqa
a_0 &=& {1 \over 2}I_{1,1},
\\
a_1 &=& {1 \over 4}I_{2,3},
\\
a_2 &=& -{1 \over 4}I_{0,1},
\\
a_4 &=& \frac{3}{16}I_{3,5},
\\
b_0 &=& {1 \over 2}I_{-1,-1},
\\
b_1 &=& -{1 \over 4}I_{0,1},
\\
b_2 &=& {1 \over 4}I_{-2,-1},
\\
b_4 &=& -{1 \over 16}I_{1,3}.
\eqa
The expressions for the remaining four
coefficients depend explicitly on $D$:
\bqa
a_3 &=& {1 \over 16 D} [5I_{5,7}-2I_{3,5}+I_{1,3}],
\label{a4}
\\
a_5 &=& {1 \over 64 D} [35I_{6,9}-10I_{4,7}+3I_{2,5}],
\\
b_3 &=& - {1 \over 16 D} [I_{3,5}+2I_{1,3}+I_{-1,1}],
\label{b3}
\\
b_5 &=& - {5 \over 64 D} [I_{4,7}+2I_{2,5}+I_{0,3}].
\label{b5}
\eqa
The factor of $D$ in the denominator arises from the angular average in 
(\ref{angular}). We have used the identity (\ref{IBP}) to put the
expressions (\ref{a4})--(\ref{b5}) into a standard form with 
no factors of $D$ in the numerator.  Setting $D=3$, we recover 
the expressions in Appendix C of Ref~\cite{Braaten-Nieto:2}.
The integrals $I_{-1,1}$ and $I_{0,3}$ 
are logarithmicly infrared divergent. With dimensional regularization, 
they have poles in $D-3$. There are therefore nonzero contributions
to $b_3$ and $b_5$ that arise from expanding the prefactor $1/D$ 
in (\ref{b3}) and (\ref{b5}) to first order in $D-3$.

There are nine additional coefficients that are required to calculate 
$\phi$, $\rho$, and ${\bf j}$  for a general time-independent state.
The expressions for these coefficients in $D$ dimensions are
\bqa
a_6 &=& -{1 \over 16D}[I_{3,5}+2I_{1,3}+I_{-1,1}],  
\label{a6}
\\
a_7 &=& {1 \over 4}I_{1,3},
\\
a_8 &=& {1 \over 8D}[5I_{4,7}-2I_{2,5}-I_{0,3}],
\label{a8}
\\
b_6 &=& {1 \over 16D}[I_{1,3}-2I_{-1,1}+5I_{-3,-1}],
\\
b_7 &=& {1 \over 4}I_{-1,1},
\\
b_8 &=& -{1 \over 8D}[I_{2,5}+2I_{0,3}-5I_{-2,1}],
\label{b8}
\\
c_1 &=& {1 \over 2D}I_{1,3},
\label{c1}
\\
c_2 &=& {3 \over 4D}I_{2,5},
\\
d_1 &=& {1 \over 2D}[I_{2,3}-I_{0,1}],
\eqa
The integrals $I_{-3,-1}$ and $I_{-2,1}$ 
are quadraticly infrared divergent, while $I_{-1,1}$ and $I_{0,3}$ 
are logarithmicly infrared divergent. With dimensional regularization, 
they have poles in $D-3$. There are therefore nonzero contributions
to $a_6$, $a_8$, $b_6$, and $b_8$ that arise from expanding 
the prefactor $1/D$ to first order in $D-3$.

\newpage
\begin{figure}[b]
\underline{FIGURE CAPTIONS:}
\caption{Feynman diagrams for calculating the coefficient $c_1$.The energy
and momentum flow counter-clockwise around the loop.}
\label{so}
\end{figure}
\setcounter{figure}{0}
\begin{figure}[htb]
\mbox{\epsfig{figure=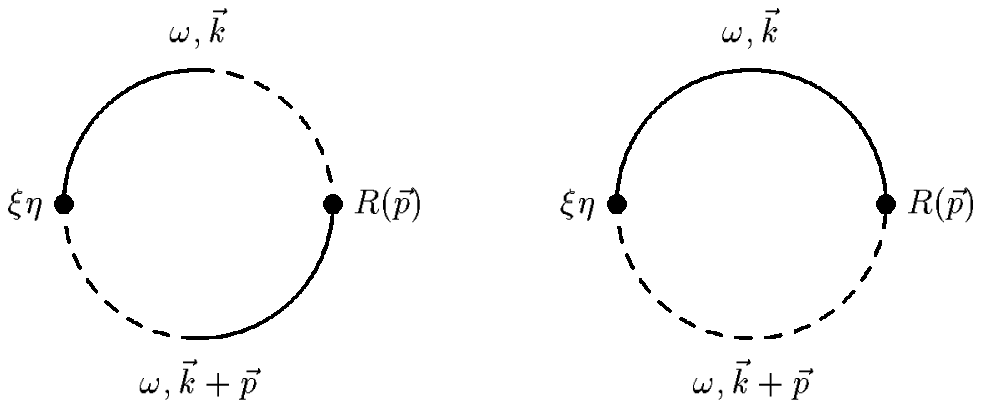}}
\caption{Feynman diagrams for calculating the coefficient $c_1$. The energy
and momentum flow counter-clockwise around the loop.}
\label{so}
\end{figure}

\end{document}